\documentclass[floatfix, aps, pra, reprint, noshowpacs, superscriptaddress,nofootinbib,longbibliography]{revtex4-2} 
\usepackage{array}[=2016-10-06]

\usepackage{algorithm}
\usepackage{algpseudocode}
\usepackage{graphicx}
\usepackage{array}
\usepackage{mathtools}
\usepackage{url}
\usepackage{hyperref}
\usepackage{natbib}
\usepackage{multirow}
\usepackage{soul}
\usepackage{xcolor}

\usepackage{mathtools,amssymb}
\usepackage{tikz}
\usepackage{tabularx}
\usepackage{enumitem}
\setlist{nolistsep}
\usepackage{censor}
\usepackage{svg}
\usepackage{wrapfig}
\DeclareUnicodeCharacter{2009}{\,}
\usepackage{hyperref}
\usepackage{caption}
\usepackage{subcaption}

\hypersetup{
    colorlinks=true, 
    linktoc=all,     
    linkcolor=blue,  
}

\newcolumntype{L}{>{\raggedright\arraybackslash}X}
\newcolumntype{C}{>{\centering\arraybackslash}X}
\newcolumntype{s}{>{\hsize=48mm}L}
\newcolumntype{b}{>{\hsize=20mm}C}
\newcolumntype{a}{>{\hsize=10.5mm}C}

\newcommand{\note}[1]{}

\begin{document}
\title{Characterizing Faculty Online Learning Community Interactions Using Social Network Analysis }

\author{Emily Bolger}
\affiliation{Department of Computational Mathematics, Science, and Engineering, Michigan State University, East Lansing, Michigan 48824, USA}

\author{Marius Nwobi}
\affiliation{Department of Computational Mathematics, Science, and Engineering, Michigan State University, East Lansing, Michigan 48824, USA}
\affiliation{Department of Biochemistry and Molecular Biology, Michigan State University, East Lansing, Michigan 48824, USA}

\author{Marcos D. Caballero}
\affiliation{Department of Computational Mathematics, Science, and Engineering, Michigan State University, East Lansing, Michigan 48824, USA}
\affiliation{Department of Physics and Astronomy and CREATE for STEM Institute, Michigan State University, East Lansing, Michigan 48824, USA}
\affiliation{Department of Physics and Center for Computing in Science Education, University of Oslo, 0315 Oslo, Norway}

\begin{abstract}

The Partnership for Integration of Computation into Undergraduate Physics (PICUP) was founded in the mid-2010s to assist educators with the challenges of integrating computation into physics curricula. In addition to in-person trainings and hosted educational materials, PICUP uses a Slack Workspace to continue collaboration and discussion offline. In this work, we use Social Network Analysis (SNA) to study the communication patterns of PICUP and assess if PICUP is meeting their goals in the Slack environment. Through our analysis, we discuss PICUP's community structure and define a conceptual framework to evaluate if the goals are being met through SNA metrics. We present a comprehensive analysis of eight channels in the Slack Workspace using various SNA metrics, identifying three distinct levels of user engagement. We conclude with implications for PICUP and provide recommendations for the community. 

\end{abstract}

\maketitle

\section{Introduction}
\label{sec:introduction}
In the past two decades, there has been a push for more students to experience and to learn computing at various stages of their career - CS 10K Project, CSforALL. These arguments range from the desire for a more computationally literate population to preparing more individuals for modern careers in science, technology, engineering, and mathematics (STEM). Related, there has been a push for integrating computation into specific disciplinary spaces to strengthen students' computing skills outside of a traditional computer science classroom \cite{silvia2019learner, magana2013introducing, lee2020computational}. This includes reports in the physics community such as the AAPT Recommendations for Computational Physics in the Undergraduate Physics Curriculum \cite{aapt} and Phys21: Preparing Physics Students for 21st-Century Careers \cite{phys21}. Compared to other disciplines, it appears that physics is leading the way \cite{chonacky2008integrating,caballero2019picup}. Regardless, instructors face numerous barriers when integrating computation into established courses, including expectations of content coverage, lack of instructor time, student resistance, and department norms \citep{leary2018difficulties, lee2020computational}.

The \href{https://www.compadre.org/PICUP/}{Partnership for Integration of Computing into Undergraduate Physics (PICUP)}  was started in the mid-2010s to assist educators with the challenges that arise from designing computing curricula for a non-computing classroom. The community supports their members in a variety of ways. From hosting in-person workshops to online webinars to example exercise sets on their webpage to a Slack workspace, instructors have a variety of opportunities to engage with material and construct a community of colleagues with similar goals. Since its inception, the community has served hundreds of faculty through their workshops and materials and has resulted in over 15 publications. 

Learning communities, particularly in the context of higher education, are nothing new. \citet{baker1999creating} describes learning communities as a group ``who have a clear sense of membership, common goals, and opportunity for extensive face-to-face interaction." \cite{cox2004building}. Often, these learning communities are situated in the Communities of Practice (CoP) framework which is defined by three components: joint enterprise, mutual engagement, and shared repertoire \cite{wenger1999communities,henderson2019faculty, tinnell2019sustaining, cox2004building}. While the definition of CoP is broad, there is an implied structure of different levels of participants, with the goal of periphery participants transitioning to core, active participants through engaging with the community. 


For faculty, learning communities can be a successful mechanism for instructor professional development (PD) when built through a sustained community of support and learning \cite{tinnell2019sustaining, furco2012using, iaquinto2011creating}. With the rapid development of technology and the ability to communicate remotely, the use of online communities as continued engagement or the sole form of PD are becoming more common as it allows educators to participate who might not be able to otherwise due to time and travel constraints \citep{henderson2019faculty,yadav2013professional, smith2012predicting, erickson2012effectiveness}. Defined by \citet{henderson2019faculty}, Faculty Online Learning Communities (FOLCs) build off existing tenets of strictly in-person Faculty Learning Communities (FLCs). The authors argue the following characteristics are some of the key components of FOLCs: periodic, potentially remote, meetings over the course of a set time; a supportive community with peers and experts; united by a common interest; foster lifelone reflective instructors \cite{cox2004building}. 


A variety of methodologies have been used to assess which strategies are most successful for sustained change in FLCs, FOLCs, and other online communities \citep{cox2004building}. Commonly, researchers conduct a pre/post survey analysis and interviews with participants about their experience, performing qualitative coding to identify trends \citep{henderson2019faculty, tinnell2019sustaining, de2019developing, furco2012using, price2021analyzing}. PICUP has done similar work with their organizers shortly after the development of the community \citep{irving2017understanding}. Similarly, researchers have used Social Network Analysis (SNA) to understand community structure and patterns in FLCs/FOLCs \citep{hatcher2022closeness, cross2006using, ma2019studying} as well as communities like Computer-Supported Collaborative Learning \citep{de2007investigating, wellman2001computer, lin2016social}. These articles utilize common metrics in SNA to describe the structure of the social interactions in these communities through identifying clusters of participants, pairs of participants, highly connected participants, less active participants, and more. SNA gives insight to the communication patterns of the community and allows us to study the structure of those patterns \citep{borgatti2018analyzing, kadushin2012understanding, newman2010networks, wasserman1994social}. \citet{dou2019practitioner}'s presents a comprehensive introduction to SNA. 

We use Social Network Analysis in this work to study the patterns of communication between members of the PICUP community on Slack. While this community is less rigid than traditional CoPs, as we will describe in Section~\ref{sec:datacontext}, PICUP's participants use Slack to discuss ideas beyond in-person meetings and thus, represents a unique, subset of the PICUP community. Similar to the works mentioned, we will use SNA metrics, such as strength centrality, closeness centrality, reciprocity, and global clustering coefficient to characterize the nature of communication in the Slack Workspace. We use these metrics to assess whether PICUP is meeting its defined goals (Section~\ref{sec:cf}). Through identifying if and how PICUP is meeting their goals in the confines of the Slack Workspace, we will be able to inform the PICUP community about changes they may want to make and motivate future analysis of the community. 


We will answer the following research questions:
\begin{itemize}
    \item What patterns of communication do we observe about the PICUP community and their Slack interactions through a Social Network Analysis lens?
    \item In what ways do these patterns align or misalign with PICUP's main goals?
    \item What changes can be made to PICUP based on this analysis? What recommendations can be made to other FOLCs when using these platforms?
\end{itemize}
Through answering these questions, we will also clearly state the limitations of our study and identify other research methods that can be used to assess PICUP's goals, beyond Slack. 

The rest of the paper is structured as follows. In Section \ref{sec:datacontext}, we provide additional context on the PICUP community and the associated Slack data. We describe the steps taken to restructure the data for network analysis and show examples for converting this data to a network structure. In Section \ref{sec:cf}, we will outline our conceptual framework by first defining various SNA metrics and tie them to PICUP's goals. In Section \ref{sec:SNA}, we construct networks and calculate the defined metrics for various channels in the PICUP Slack Workspace. Section \ref{sec:implications} describes the results of our analysis and the implications for the leaders of PICUP, and Section \ref{sec:limitations} states the limitations in our analysis and related claims.

\section{Data Context and Preparation}
\label{sec:datacontext}
In this section, we provide additional context on the structure of the PICUP community, particularly their use of Slack. We describe our data preparation process of converting the raw Slack messages into a SNA friendly format. Then, we show the step-by-step process of representing the Slack messages as a social network while describing the various choices for representing communication in a network. Our detailed transparency in this process is to highlight what information is captured by these networks (and subsequently, what is not). These network representations can show a condensed visualization of the communication in the Slack channel, but are not the sole mechanism for extracting information.

\subsection{PICUP Community Structure}\label{sec:picupcs} 
The creators of PICUP developed the group as a space for instructors to learn and communicate with one another about pedagogical tools for integrating computation into physics curricula. 
In the early days of PICUP, \citet{irving2017understanding} demonstrated the community had potential to meet the framework of the traditional CoPs and FOLCs. While PICUP meets many of the definitions for CoPs and FOLCs, it has a less rigid structure than traditional settings. PICUP offers workshops, meetings, and meetups at conferences throughout the year, but these are not required meetings for the group. The group is ongoing and anyone can join at any time. There are resources shared with the community and workshops for helping educators implement them, but instructors can also pull directly from the site and implement it on their own. They are united by the common interest of integrating computing into their physics-focused classrooms, but there is a wide depth and breadth in the types of instructors that engage with PICUP's materials. PICUP offers a supportive community to engage and learn from experts, but this happens through formal means like the workshops as well as informal means like the Slack Workspace. Following the traditional CoP framework, there are opportunities for instructors to move from the periphery to become more active, expert-like members of the community, but this is dependent on the engagement and interest of the individual and not inherent to the community. 

With this in mind, PICUP creates a community of researchers with likeminded goals, but does not fit into the traditional, rigid definitions of Communities of Practice and Faculty Online Learning Communities. This paper does not seek to define what PICUP currently is, evaluate further what components of FOLCs it meets, identify if subgroups of PICUP follow CoP, or if it has the potential to be classified as a traditional FOLC. The project has been funded by another grant, which will seek to answer some of these questions. 

The Slack Workspace also does not follow a traditional CoP or FOLC framework. Rather, it is a component of PICUP and is a mechanism for further communication. While it may help enhance PICUP's FOLC status and aid its members in becoming expert level, this is not something we can evaluate directly without analysis of the entire PICUP community and support structure. Additionally, the Slack community only represents certain users and the experiences of those users in a particular environment. While SNA will give insight on communication patterns, they are restricted to the community interactions on Slack. Because PICUP's Slack community cannot be mapped directly to CoP, we have defined our own conceptual framework connecting the community, SNA metrics, and PICUP's goals (Section~\ref{sec:cf}). Our analysis in this work will additionally justify why PICUP does not meet the traditional definitions for FOLCs or COPs.

\subsection{From Slack Messages to Processed Data}
\label{sec:background}

While SNA representation does not retain the content of the Slack messages, we take several steps to ensure as much of the communication is retained as possible. In this subsection, we describe the process to restructure the Slack messages into a usable format for SNA.

Slack (Searchable Log of All Communication and Knowledge) is an instant messaging platform used by many research teams and businesses \cite{Slack}. A workspace hosts a community for users in a specific team or company. Within the workspace, users can create channels for communication dedicated to certain topics. Channels can be made public or private. Lastly, users can message another or multiple users directly - outside the channels. A user can be a part of multiple workspaces and multiple channels within that workspace. 

While the PICUP Slack Workspace is still open, our data contains messages from August 2016 to June 2021. The total number of messages sent in the entire workspace prior to any data reformatting is 9,139. Our work only includes messages sent in public channels, not messages sent in private channels or direct messages. (We do not included messages sent in private channels as we do not have access to that data.) The total number of users at the end of data collection was 475. When a user joins any Slack Workspace, they are automatically added to the `General' and `Random' channels. Users must opt-in or be invited to other channels in the network. 

In Slack, a user may use tags to notify specific users in the channel. For example, a message containing `@user1' is visible to everyone in the channel, but only notifies user1 about the given message. Messages that tag a specific user only result in one entry to the dataset per user. For example, if user2 sends the message, \textsl{``@user3 can you please send me the textbook you are using for your intro to physics class?"}, our data would contain an entry only from user2 to user3 with the associated message content and timestamp. 

Users can also send a message to all the others users in a channel by using `@channel' in the given message. Each `@channel' message is translated into multiple entries between the sender and all current users in the channel. For example, if user1 send a message, \textsl{``Hi @channel, I have a question about Python!"}, our dataset would contain a single entry from user1 to every current user in the channel with the same timestamp and same message. All messages that do not contain a tag to a specific person, even if they do not include an `@channel' tag, are treated in this manner. To avoid alerting the whole channel with a message in response to another message, a user can ``reply in thread.'' This creates a dialogue between the sender and any other user who chooses to message in the thread. As we did before, we create a new entry between the sender and the current users in the thread with the message sent. 

In our dataset, the last column `To\_Channel' indicates whether the message was sent to the entire channel (True) or directed at a specific person (False). For messages tagged for a specific person, the `Userid\_To' column contains the user that was tagged. For messages that were sent to the whole channel, the row value is repeated for the number of total users in the channel with the `Userid\_To' column is individually filled with each individual user id. 

\begin{figure}[htp]
    \centering
    \includegraphics[width=.45\textwidth]{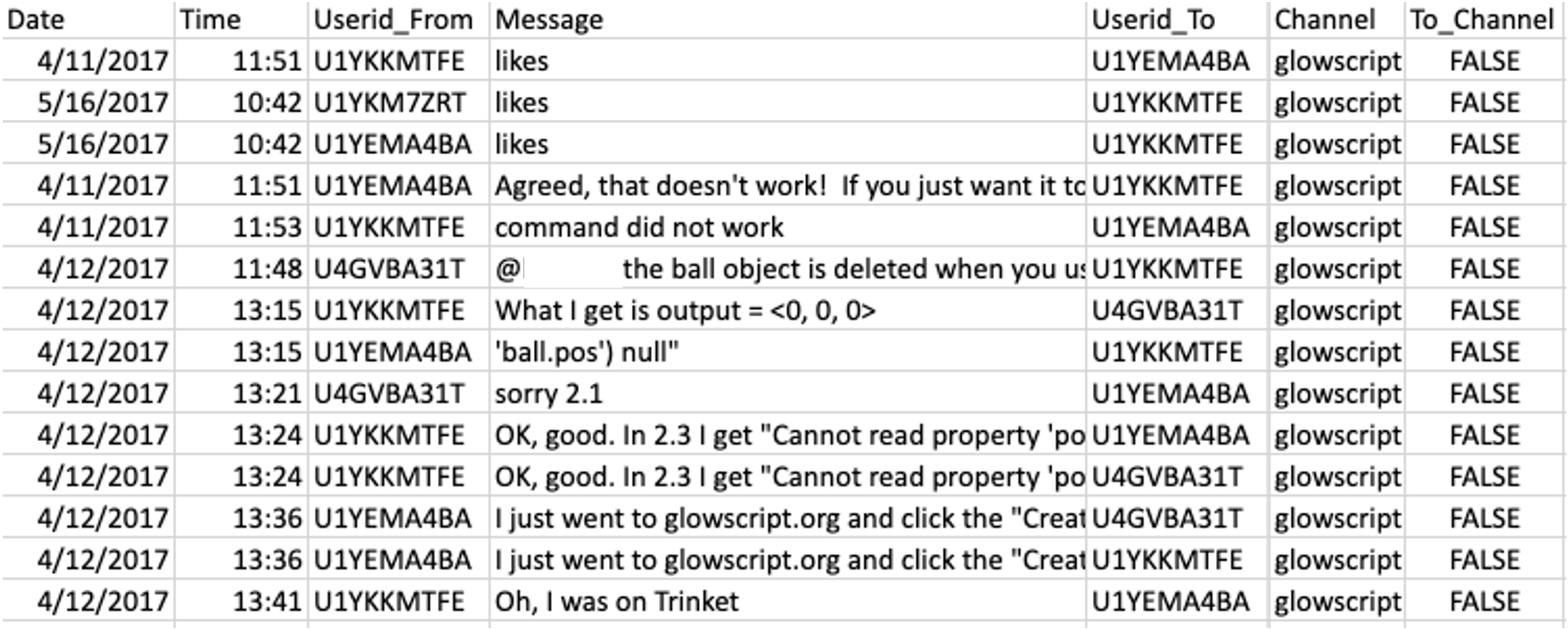}
    \caption{A snapshot of the restructured data from the Glowscript channel. The figure displays the post-processed data following our steps to identify receivers and senders, transform messages sent to the entire channel, and represent emoji responses as messages. }
    \label{fig:data}
\end{figure}

Additionally, any user may respond to a message with one or multiple emojis. Because the emoji reactions are still a form of communication between two users, it is crucial to retain this information in the network representation. Thus, we translate each emoji reaction into a message between the two users. For example, if a user reacted with a \href{https://emoji.slack-edge.com/T02291UPX55/vibepartycat/2c7317181cfb7286.gif}{``:vibepartycat" emoji}, we generate a message from the receiver to the sender with the emoji description as the content. Other smaller cleaning processes include: adding the file name to the message where files were attached, adding communication type where possible (e.g. join/leaves), and removing bots from users and messages. Figure~\ref{fig:data} shows a snapshot of this cleaned data. 

At the end of the data processing stage, we separate the data by channel to facilitate network building. There are 44 total channels in the Slack Workspace. Only 13 of these channels have more than 40 messages prior to data restructuring. In Table~\ref{tab:messages_compare}, we select a few channels to compare the number of messages in the pre-processed and post-processed data to highlight the variance in activity between the channels. The channels are presented by increasing total number of users in the Slack channel, with Advanced Thermodynamics being the smallest and Random being the largest. The associated networks for each of the channels are in Section~\ref{sec:SNA} and in the Appendix \ref{appendix1}. For networks of similar size (in both number of pre-processed entries and number of users), we see similar trends to the number of post-processed entries. For example, there is about a 12x increase in the number of messages between pre- and post-processing for the Advanced Thermodynamics and Trinket channels. These channels also have about the same number of users, 30 and 37 respectively. 

The number of messsages in the post-processed data is heavily dependent on the number of users in the channel and the number of messages sent to the entire channel. Each message sent to the entire channel contains an entry to every other user in the channel. Thus, for a large number of users and even a moderate number of messages sent to the entire channel, the total number of messages increases dramatically. Comparing the Jupyter and Random channels, while they have a similar amount of messages prior to data processing, the Random channel has nearly 5.5x more entries post data processing, due to the channel containing about 3.5x more users. These type of messages contribute most to the drastic increase in total number of messages compared to messages tagging a user directly or an emoji reaction. 

We note that the total number of users in Table~\ref{tab:messages_compare} might different from the total number of users seen in the Figures below. Slack generates a messages each time a user joins and leaves the channel. While these messages are in our data set, we do not use them to construct the networks as they cloud the main interactions between users. If users join at the end of data collection and there are no additional messages following their join message, they will appear in our total count in number of users however, they will not appear in our network representations as they have not been a part of any interactions in the network yet. The number of users with at least one message in the network can be seen in the last column of the Table. 

\begin{table*}[t]
    \begin{center}
    \caption{The table shows the channel name, number of raw messages (i.e. distinct messages sent in Slack), number of messages post data restructuring (i.e. the total number of interactions), the total number of users, and the number of users that have sent or received at least one message. As seen, the number of users, and the number of messages sent to the entire channel highly affects the number of messages in the post data restructuring column. }
    \label{tab:messages_compare}
    \begin{tabularx}{0.85\linewidth}{c|c|c|c|c}\hline\hline
        Channel Name & Raw Data & Total Interactions & Total Number Users & Users with $\geq$ 1 Message \\\hline
        Advanced Thermodynamics & 60 & 678 & 30 & 16\\
        Advanced Mechanics & 49 & 770 & 34 & 24\\
        Why We Do This & 83 & 1224 & 35 & 23\\
        Upper Mid-West & 247 & 3582 & 31 & 30\\
        Classroom Pedagogy & 102 & 3,338 & 72 & 58\\
        Trinket & 133 & 1,839 & 38 & 37\\
        Glowscript & 476 & 17,618 & 92 & 91\\
        Jupyter & 776 & 28,926 & 128 & 120 \\
        Random & 966 & 163,296 & 459 & 457\\\hline\hline
    \end{tabularx}
\end{center}
\end{table*}

\subsection{From Processed Data to Networks} \label{sec:dattonet}
\begin{figure}[htp]
    \centering
    \includegraphics[width=.45\textwidth]{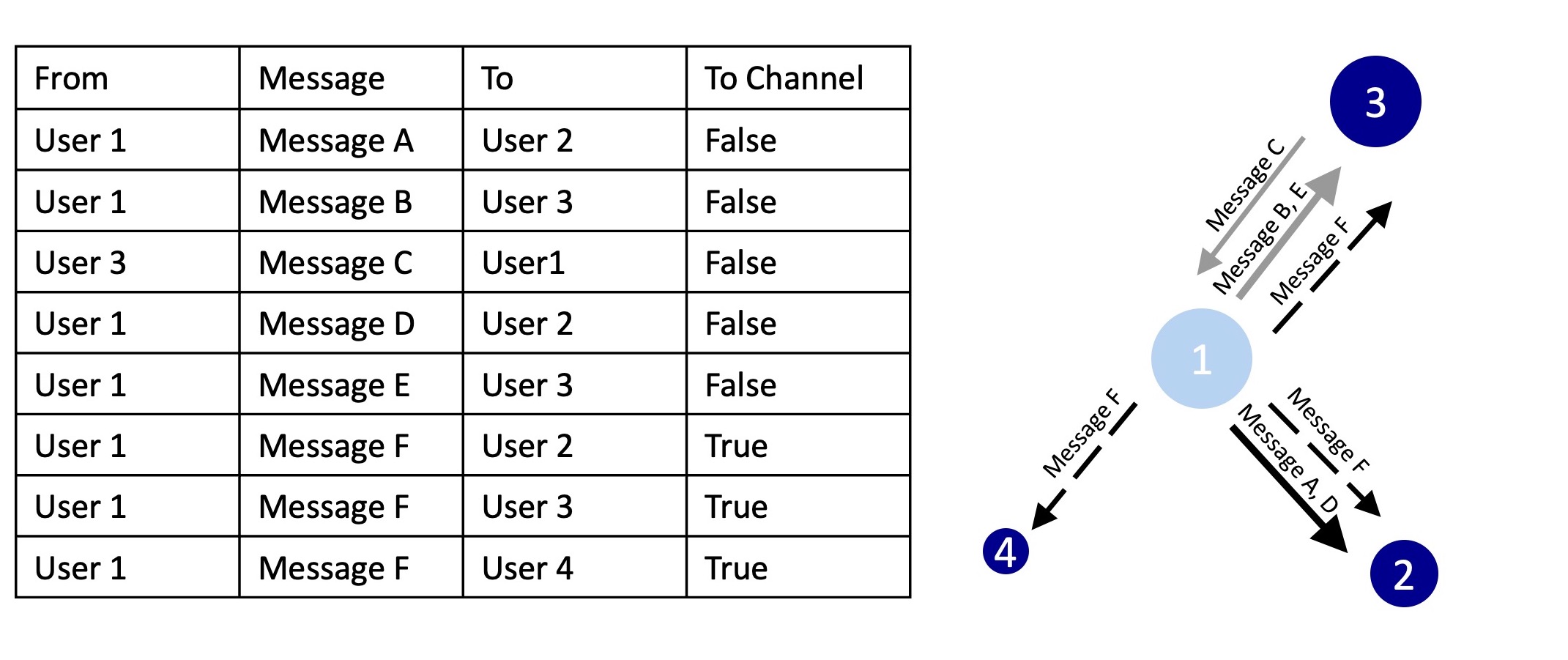}
    \caption{We present an example to show how the structured data is represented as a network in our project. Users are represented with nodes, messages are represented with directed edges, using size and color to indicate user attributes. In our representation, solid edges indicate direct messages between two users. Opposingly, dashed edges indicate a message that was sent from one user to the entire channel. Edge width corresponds to the number of messages between two users. }
    \label{fig:example_network}
\end{figure}

In Figure~\ref{fig:example_network}, we illustrate how networks are generated from post-processed data. The table is an example post-processed data and the right is the associated network representation. We are using \textit{directed and weighted} networks. Each user is represented with a node and the corresponding number. Messages are represented with directed edges to indicate the sender and receiver. For example, Message A is represented with an edge from User 1 pointing to User 2. 

If multiple messages are sent between the same two users, we increase the weight of the edge. Message D is an example of this. Notice, the solid edge from User 1 to User 2 is thicker than the solid edge from User 3 to User 1 from Message C. We use edge type to denote the type of message. Each of these first five messages were directed at a specific user and hence, the associated edge was a solid line. The remaining rows in the table represent a message sent from User 1 to every other user in the channel - represented with a dashed edge. 

The edge color indicates whether there is a reciprocated edge between two users. In our example, User 1 has sent Message B and Message E to User 3, and User 3 has sent Message C to User 1. Because both users have sent a message to the other, the edge is colored grey. Opposingly, User 2 has not sent a message back to User 1 and hence, the edge is colored black. As we will see in the more complicated networks, this simple coloring can help us contextualize the network quickly.

The node size and node color are used to represent node strength. We present a formal definition in Section \ref{sec:SNAmetrics}. Put simply, strength is the sum of the weights of the edges associated to a specific node. The in-strength of a node is the sum of the weights of the edges pointing towards the node and the out-strength is the sum of the weights of the edges leaving the node. 

While we visually separate the edges sent directly to another user and to the entire channel, they are combined in the strength calculation. In Figure~\ref{fig:example_network}, Node 3 has a total strength of $4$ with an in-strength of $3$ and an out-strength of $1$. Nodes with a larger out-strength than in-strength (sending more messages out than receiving) are colored light blue, such as Node $1$. Users with a larger in-strength are colored dark blue, such as Nodes $2,3,4$.

\section{Conceptual Framework}
\label{sec:cf}
In this section, we define our conceptual framework for mapping PICUP's goals to Social Network Analysis metrics. To define this framework, we need to introduce and define the SNA metrics we use in this work. Then, we will present our conceptual framework by connecting PICUP's goals with the associated expectation in the network structure and the metrics we will use to assess these goals.   

\subsection{Social Network Analysis Metrics}\label{sec:SNAmetrics}
Network science has a variety of metrics to characterize the nature of communication in the network \cite{kadushin2012understanding,borgatti2018analyzing, newman2010networks}. Most metrics were initially constructed for unweighted and undirected network. However, with weighted and directed networks, it is critical to incorporate these aspects into the metric calculation as it can affect the value, and thus the interpretation, of the metric \citep{dou2019practitioner,kadushin2012understanding}. Often, there are many options for a single metric that account for directionality and weighted edges of a network, which allows researchers to pick a variation that fits what these features represent in their networks \citep{opsahl2009clustering, fardet2021weighted, schank2005approximating, liu2014weighted, wang2015review}. We will discuss the ones we have chosen for our networks in this section. All analysis was conducted in R \citep{R} with specific packages cited below. 

\subsubsection{Strength}
As described in Section~\ref{sec:dattonet}, the strength of a node, $i$, is the sum of the weights of edges associated that node, 
\begin{equation*}
    s_i = \sum_{i \neq j} w_{ij},
\end{equation*}
where $w_{ij}$ is the edge weight between nodes $i$ and $j$ \citep{borgatti2018analyzing, kadushin2012understanding, fornito2016fundamentals}. 

The in-strength is
\begin{equation*}
    s_i^{\text{in}} = \sum_{i \neq j} w_{ij}^{\text{in}},
\end{equation*}
where $w_{ij}^{\text{in}}$ is the weight of the directed edge to node $i$ from node $j$, and the out-strength is 
\begin{equation*}
    s_i^{\text{out}} = \sum_{i \neq j} w_{ij}^{\text{out}},
\end{equation*}
where $w_{ij}^{\text{out}}$ is the weight of the directed edge from node $i$ to node $j$  \citep{borgatti2018analyzing, fornito2016fundamentals}. We use igraph's built-in functions for these calculations \citep{igraph1, igraph2} . 

\subsubsection{Closeness Centrality}
While the strength metric identifies frequent senders and receivers of Slack messages, we calculate closeness centrality to identify the central users in the network \citep{bavelas1950communication, freeman2002centrality}. In an undirected, unweighted network, a node's closeness centrality is the average of the shortest paths from the node to every other node \citep{borgatti2018analyzing, opsahl2010nodecentrality}. 

To include edge weight, we must adjust the distance measurement between two nodes, and the simplest extension of the shortest path would be the smallest total distance between the nodes. However, this prioritizes paths with smaller edge weights which is not always the correct assumption for the application. In some cases, like in our networks, a stronger edge weight between two users can represent a stronger connection, and thus, it should represent an easier path to traverse through. A user with multiple edges with large weights should have higher closeness centrality value than a user with mutliple edges with low weights. See \citet{opsahl2010nodecentrality} for specific examples. 

Following \citet{opsahl2010nodecentrality}, we update the distance metric between to nodes $i$ and $j$ for intermediary nodes $h$ to be:
\begin{equation*}
    d_{w\alpha}(i,j) = \text{min} \left (\frac{1}{(w_{ih})^\alpha} + ... + \frac{1}{(w_{hj})^\alpha}\right ), 
\end{equation*} 
where we simply invert the edge weights between two nodes. Additionally, we incorporate a tuning parameter, $\alpha \in (0,1)$, to balance the effect of the weights of the intermediary nodes. For smaller values of $\alpha$, the longest distance is the path with the greatest number of intermediary nodes. For larger values of $\alpha$, the number of intermediary nodes is not weighted as heavily compared to the strength of the edges. 

With the inversion of the edge weights, the closeness centrality (CC) becomes:
\begin{equation*}
    CC_{w\alpha}(i) = \left[ \sum_j^N d_{w\alpha}(i,j)\right]^{-1},
\end{equation*}
and the interpretation as expected is preserved.

This metric can easily be extended to account for directed edges. For each node, we calculate a closeness centrality out (CC\_{out}) - available paths only include edges leaving the node - and a closeness centrality in (CC\_{in}) - available paths only include edges coming towards the node. Users with a large CC\_{out} are good broadcasters of information. They are central in the network and have short, heavily weighted paths to many other users in the network. Users with a larger CC\_{in} are central in receiving information. Using the directed and weighted implementation, users can fall into one of the four categories displayed in Table~\ref{tab:cc_compare}.

\begin{table}[t]
    \centering
    \caption{The four classifications of closeness centrality values. Generally, users with a larger out closeness centrality have short outward paths to other users in the network thus, are good broadcasters of information. Opposingly, users with a large in closeness centrality are frequent receivers of information.}
    \label{tab:cc_compare}
    \begin{tabularx}{\linewidth}{c|c}\hline\hline
        CC Value & Interpretation \\\hline
        Large CC\_{out}, Large CC\_{in} & Good Broadcaster\\
                                        & Frequent Message Receiver\\
        \hline
        Large CC\_{out}, Small CC\_{in} & Good Broadcaster \\
            & Few Messages Received \\
        \hline
        Small CC\_{out}, Large CC\_{in} & Poor Broadcaster\\
            & Frequent Message Receiver \\
        \hline
        Small CC\_{in}, Small CC\_{in} & Poor Broadcaster\\
            & Few Messages Received \\
        \hline
        \hline\hline
    \end{tabularx}
\end{table}
Similarly to strength, these classifications will assess whether PICUP is meeting its goals by identifying active users in the network and their associated influence. We will likely see the users spread across the four categories, but ideally in a well-connected network, many nodes would have a similar score. 

We calculate the closeness centrality metric using \citet{tnet}'s `closeness\_w' function. 

\subsubsection{Weighted Reciprocity}
In addition to highlighting reciprocal communication with grey edges in the networks, we calculate reciprocity to understand the total amount of reciprocal communication in our networks. Reciprocity is the probability that the opposing counterpart of a directed edge is also included in the edge set \citep{newman2010networks}. For undirected and unweighted networks, it is calculated as number of reciprocated edges divided by the total number of edges \citep{borgatti2018analyzing}. This metric is bounded between 0 and 1, where networks with a larger reciprocity have a larger probability of reciprocated ties. 

However, this implementation of reciprocity does not account for edge weight - just whether there exists a reciprocated edge between two users. To represent edge weight appropriately in the metric, we follow \citet{squartini2013reciprocity} and define weighted reciprocity as 
\begin{equation*}
    w_r = \frac{\overleftrightarrow{W}}{W} = \frac{\sum_i \sum_{j\neq i} \overleftrightarrow{w_{ij}}}{\sum_i \sum_{j\neq i} w_{ij}},
\end{equation*}
where 
\begin{equation*}
\overleftrightarrow{w_{ij}} = \min\{w _{ij}, w_{ji}\} = \overleftrightarrow{w_{ji}}
\end{equation*}
for pairs of nodes $i,j$.

More plainly, we obtain a single value for the denominator by summing all the weights in the network. For the numerator, we generate a variation of the network to represent each interaction equally. If there is an edge from User 1 to User 2 with a weight of 3 and an edge from User 2 to User 1 with a weight of 5, the weight of both edges becomes 3. We sum these weights from this new network to obtain the numerator value. This weighted version is analogous to the original metric, while accounting for the weights of the edges. This extension is still bounded between $0$ and $1$. 

We will use this metric to assess PICUP's goal of creating and growing community in the Slack Workspace. High amounts of reciprocity, values closer to 1, indicate heavily reciprocated communication between the users and imply there are relationships at least amongst pairs of users in the network. Ideally, to emulate a strong online community, we want users to be active and responsive to their peers. This metric does not identify the type of communication, but gives insight to how much commmunication is happening in the network. 

\subsubsection{Weighted Global Clustering Coefficient} 
Clustering Coefficients are another common metric for characterizing whole network structure. To identify areas of high and low density in a network, \citet{watts1998collective} proposed the local clustering coefficient which measures the likelihood the neighbors of a particular node are also connected. The local clustering coefficient of a node, $v_i$, in an undirected graph is the number of pairs of neighbors of $v_i$ that are connected divided by the number of possible connections between pairs of neighbors of $v_i$, where neighborhood is the sub-network that contains all the degree $1$ connections of $v_i$ \citep{borgatti2018analyzing, wasserman1994social,opsahl2009clustering}. 

\begin{figure}[h]
    \centering
    \includegraphics[width=.45\textwidth]{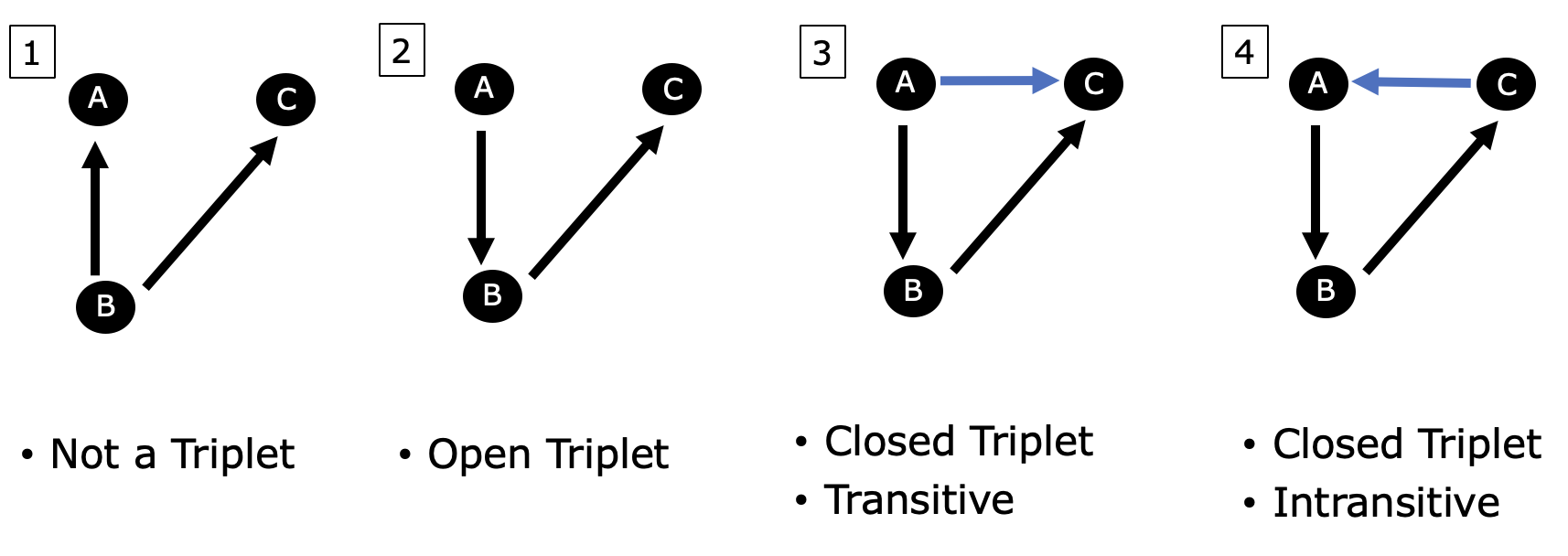}
    \caption{Differentiating transitive and intransitive triplets. Triplets defined by \citet{wasserman1994social} are a 2-path - there is a path from node A to B and node B to C as seen in (2). A transitive triplet follows the transitive property implying that if there is a triplet from nodes A to B and B to C, the triplet is transitive if there exists a path from A to C (as seen in (3))\citep{opsahl2009clustering}.}
    \label{fig:triplets_dir}
\end{figure}

To extend to a network-wide metric, we average the local clustering coefficient across all nodes in the network \citep{watts1998collective,schank2005approximating}. The global clustering coefficient (GCC) can be equivalently defined as
\begin{equation}\label{eq:GCC_triplet}
    \frac{\text{number of closed triplets}}{\text{number of open triplets} + \text{number of closed triplets}},
\end{equation}
where closed triplets are a set of three nodes and three edges with one connection between each of them, and open triplets are a set of three nodes and two edges \citep{opsahl2009clustering, newman2003structure}. See (2) and (3) in Figure~\ref{fig:triplets_dir} for an example of an open and closed triplet.

\begin{figure}[h]
      \centering
    \includegraphics[scale=0.35]{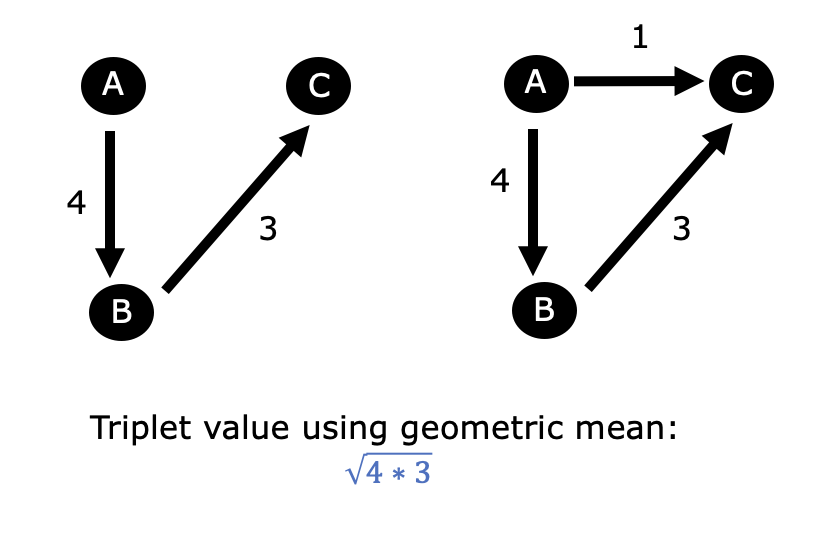}
    \caption{For the weighted and directed GCC, we calculate a triplet value for each counted triplet in the metric. We use the geometric mean as shown. The geometric mean for an open and closed triplet with edge weights of $3$ and $4$ is $\sqrt{4*3}$}. 
    \label{fig:triplet_value}
\end{figure}

By default, the global clustering coefficient above does not account for directed or weighted ties. \citet{opsahl2009clustering} account for this in their version of the metric. For directed networks, our equation for GCC still follows Eq.~\ref{eq:GCC_triplet}, however we change which triplets in the network get counted. In a directed network, an open triplet at node B occurs at (2) in Figure~\ref{fig:triplets_dir} - the directed edges must pass through B. Further, a closed triplet can now be transitive or intransitive. As with a traditional definition of transitivity (if A is connected to B and B is connected to C, A must be connected to C), we call (3) in Figure~\ref{fig:triplets_dir} a transitive, closed triplet. In this metric, we only count qualifying open triplets and closed, transitive triplets. To account for weighted edges, we continue to build off the previous definition by attaching a triplet value to each open and closed triplet. Eq.\ref{eq:GCC_triplet} becomes    
\begin{equation}\label{eq:GCC_triplet_w} 
    \text{WGCC} =
  \frac{\text{total value of closed triplets}}{\splitfrac{\text{total value of open triplets}}{ + \text{total value of closed triplets}}}.
\end{equation}
There are many options for the triplet value, we use the geometric mean as seen in Figure~\ref{fig:triplet_value}. This metric is more robust against differences in unbalanced weights and it bounded between $0$ and $1$. 

We will use this value to asses PICUP's goal of continual community growth and strength in community building. WGCC values closer to $1$ indicate many closed transitive triplets and thus, indicate a high  connectedness amongst the community members. Ideally, in PICUP's networks, we want this value to be larger to indicate high amounts of communication between groups of three in the community. 

We use \citet{tnet}'s `Clustering\_w' function for this calculation. 

\subsection{Random Networks and the Configuration Model} \label{sec:config}

For the whole network metric calculations, we would like to study how (non)unique the value is compared to networks of similar size and distribution. This falls under the umbrella of inferential statistics, where we analye the likelihood of observing a phenomena about a sample from a population by chance. 

In traditional hypothesis testing, we compare an observed result in the sample to some null value, where we are looking to identify if the sample result differs from what is assumed under the null hypothesis. However, the approximations needed to calculate these tests are only valid when the observations are independent - an assumption which does not generally hold true for social network data \citep{hanneman2005introduction,hobson2021guide,croft2011hypothesis}. Specifically, we cannot generally assume that the individuals in the network do not influence each other's actions \citep{cranmer2017navigating}. Additionally, while it would be ideal, we have no theoretical baseline distribution to compare our network metric value to. For example, we do not know the distribution of values for weighted reciprocity for networks with the same number of nodes and edges as our original network.

To statistically quantify our metric value, similar to a traditional z-test, we need to generate our own sample of similar networks to the reference (or original) network \citep{hobson2021guide}. In this method, we are essentially estimating the probability distribution directly. Following the suggestion of \citep{hanneman2005introduction, maslov2004detection}, after generating an ensemble, we calculate the metric we are interested in for the original network and each of the networks in the ensemble. We then calculate the proportion of random networks in the ensemble with a lower metric value than the original network. Extreme probability values indicate that our network value is different than those in the sample (similar to a traditional confidence interval). Opposingly, probability values closer to 0.50 indicate our network metric is not different than networks of similar size, shape, and distribution. This value is an estimator of the likelihood that the metric value of the original network occurs by chance. With this method, we assume the generation technique creates a random enough ensemble such that the networks in the ensemble can be assumed to be independent, and thus, we can assume our probability value is reasonably accurate. For more details on this type of statistical analysis, please see \citet{marion2016evaluating}. 

\begin{figure}[htp]
    \centering
    \begin{subfigure}[b]{0.38\textwidth}
        \centering
         \includegraphics[width=\textwidth]{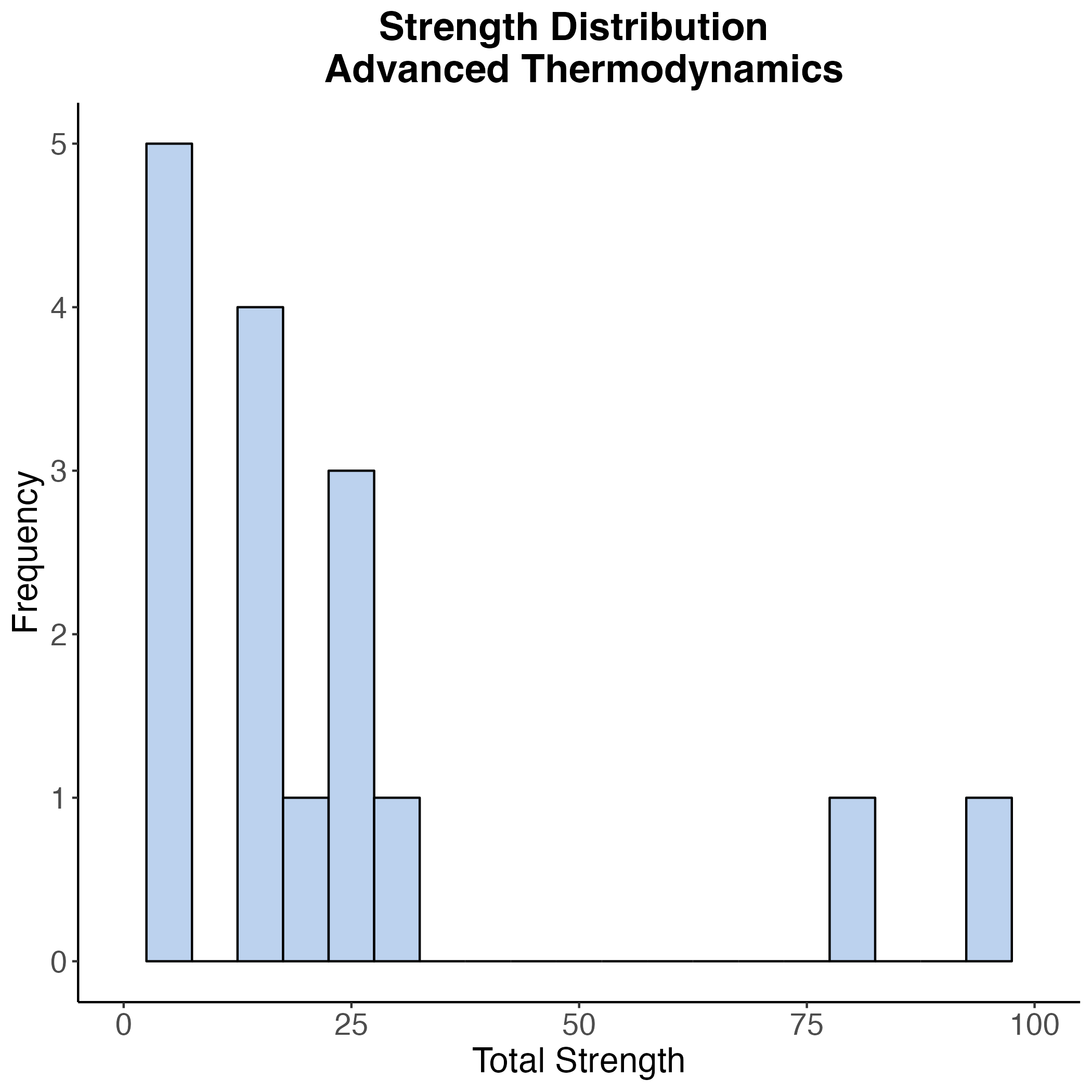}
        \caption{} 
        \label{fig:AT_strength}
        \end{subfigure}
    \begin{subfigure}[b]{0.38\textwidth}
       \centering
       \includegraphics[width=\textwidth]{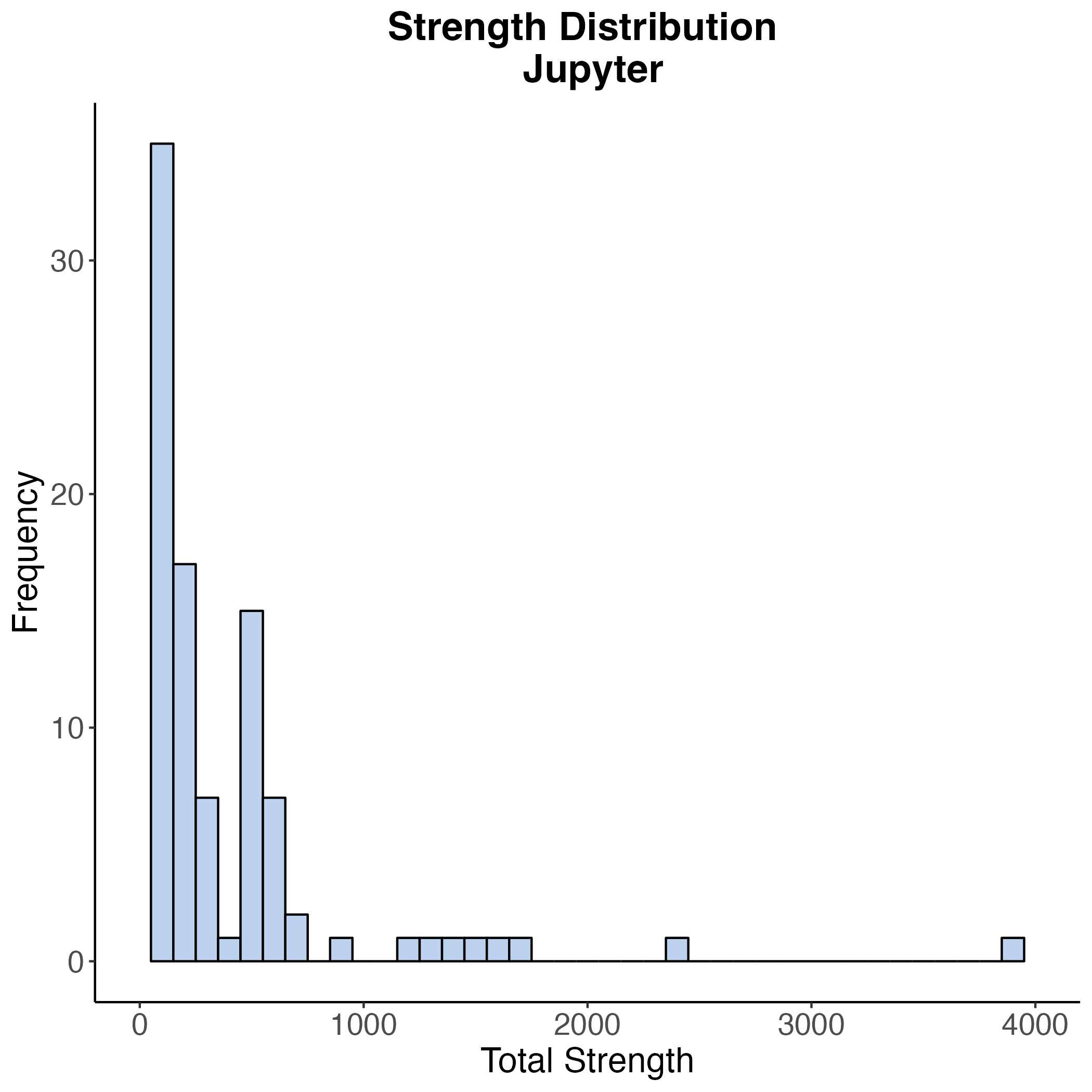}
        \caption{ } 
        \label{fig:J_strength}
    \end{subfigure}
    \begin{subfigure}[b]{0.38\textwidth}
       \centering
      \includegraphics[width=\textwidth]{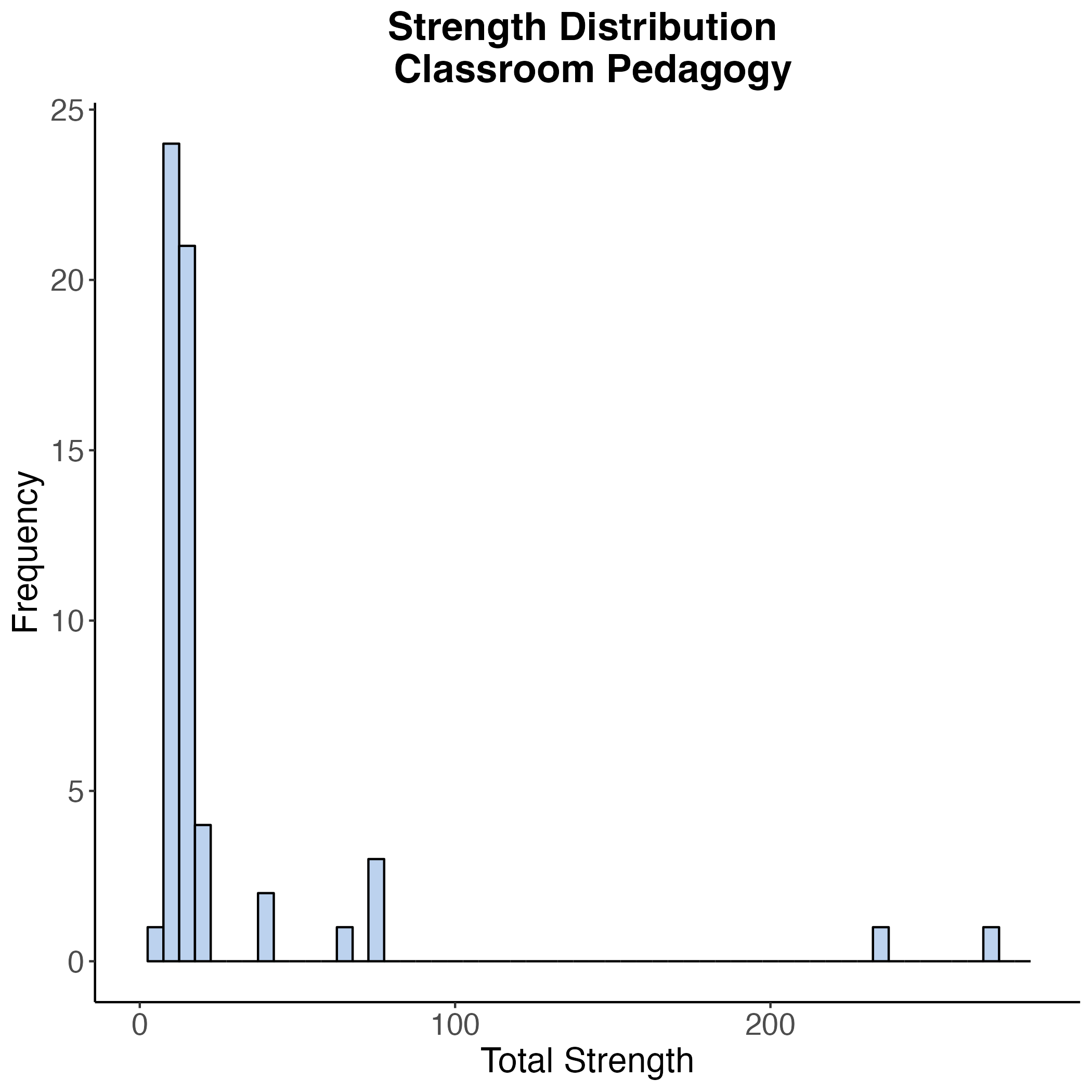}
        \caption{ } 
        \label{fig:CP_strength}
    \end{subfigure}
       \caption{The strength distributions for the main networks analyzed in Section~\ref{sec:SNA} - Figure~\ref{fig:AT_strength} for the Advanced Thermodynamics,  Figure~\ref{fig:J_strength} for the Jupyter, Figure~\ref{fig:CP_strength} for the Classroom Pedagogy.}
       \label{fig:all_strength}
 \end{figure}

Often, we use a strength distribution to characterize a network \citep{hobson2021guide, croft2011hypothesis, fornito2016fundamentals,newman2008physics}. As previously defined, a node's strength is the sum of the weights of the edges connected to the node. We use a histogram to represent the distribution of strength values for nodes in the network. Figure~\ref{fig:all_strength} shows the strength distributions for all three networks we will perform an in-depth analysis on in Section~\ref{sec:SNA}. The strength distributions are heavily right skewed. Often, these types of networks are called scale-free networks \citep{newman2010networks}. The strength distribution is a key choice in the method choice for generating new, similar networks as it helps assess similarity between the original network and the networks in the ensemble. Additionally, the choice of network generation method can lead to false conclusions, especially with smaller amounts of data \citep{hobson2021guide}.

With this in mind, to create our random ensemble, we implement the Configuration Model \citep{newman2001random, newman2003structure, newman2010networks}. The method preserves the strength distribution of the reference network as our distribution to generate random networks. In its simplest form, the configuration model can be thought of as a probabilistic reshuffling of the edges in the network. We fix the in and out-strength for each node from the original network as well as fix the number of edges. At each iteration, we pick a head and tail node with probability inversely proportional to the sum of the heads (tails). An edge is added or the weight is updated between the head and tail nodes accordingly. For a walkthrough example with a smaller network, see Appendix~\ref{appendix2}.

When implementing the configuration model, there is often a concern for repeated configurations in the generated ensemble. This is particularly true for small networks with few edges. Additionally, while the matchings of heads and tails occur with equal probability, the full network configurations do not occur with equal probability. To address this concern, we ran simulation test to quantify the amount of repeated, identical networks in the ensemble of random networks. Due to the complexity of our networks, we are not concerned with repeated, identical networks. See Appendix~\ref{appendix3} for more information. 

For the Advanced Thermodynamics and Classroom Pedagogy channels, there are 7000 networks in our ensemble. For the Jupyter channel, there are 3000 networks in the ensemble. These numbers were chosen to balance computation time as well as statistical accuracy, while minimizing the number of identical repeats.

\subsection{An Ideal Network Structure: Our Conceptual Framework}
In this work, we are assessing three of PICUP's goals \citep{caballero2019picup}:
\begin{enumerate}
    \item Lowering barriers for faculty to integrate computation into courses
    \item Continual community growth and strength in community building
    \item Increased growth in number of community leaders
\end{enumerate}
Ideally, continually meeting these goals will encourage more faculty to interact and engage with the community. In Table~\ref{tab:cf}, we define our framework connecting PICUP's goals, the SNA metrics to assess these goals, and the metric values we expect if the goals are being met.

\begin{table*}[htp]
        \centering
        \caption{The Conceptual Framework mapping three of PICUP's goals to expectations of community structure and metrics from Social Network Analysis to assess whether the goals are being met in the Slack Community.  }
        \label{tab:cf}
        \begin{tabularx}{\linewidth}{p{5cm}|p{5cm}|p{5cm}}\hline\hline
            PICUP Goal & Expectation & SNA Metric  \\\hline\hline  
            Lowering barriers integrate computation into courses & Many Active and Types of Participants & (High) Strength, (Low) in-strength, (High) out-strength    \\
            \hline
            Continual community growth & Reciprocated Communication, Many Active Users, Many Clusters & (High for all) Weighted Reciprocity, Closeness Centrality, Strength, Weighted Global Clustering Coefficient \\
            \hline
            Increased number of community leaders & Active Users Beyond Leadership &  (High) Closeness Centrality, (High) Strength Across Different Levels \\ \hline\hline
        \end{tabularx}
    \end{table*}

 PICUP's main goal is to lower barriers for faculty to integrate computing into their classrooms. We expect the Slack community to aid in this endeavor by creating an accessible mechanism for communicating with others in the community. With low barriers, users in the network would be comfortable communicating with others in the networks and sharing their ideas. To meet this goal, we would expect the networks to have many participants, many active participants, and many types of paricipants. This would imply there is a lot of communication happening in the community, and the community is conducive to asking questions or sharing ideas. In the networks, we can assess this goal through metrics like strength, in-strength, out-strength. Ideally, there would be a lot of users with large strength values and out-strength values greater than or comparable to in-strength. While ideal, this is not always expected for a traditional social network \citep{sun2014understanding}. With that in mind, we would like this to happen with a large subset of the users in the network and have few users on the periphery of the network with large in-strength values. 

Another goal of PICUP is to continue to grow the number of members engaging with the materials and create a strong sense of community. Again, the Slack should provide a remote oppportunity for members to communicate and discuss ideas offline, and be an easy mechanism for engagement with others in the group. To meet this goal, we would expect high values for the whole network metrics - weighted reciprocity and weighted global clustering coeffficent - particularly in relation to networks of similar structure. High values would indicate much reciprocal communication and small clusters of users frequently communicating. Additionally, we would expect high values for closeness centrality and strength values. These would help identify the strength in the community structure in each of the channels, and assess community growth.  

PICUP is led by a group of experts in the computational physics field. However, building off the previous two goals, PICUP seeks to use their community to create more leaders and experts in their field. This is slightly more challenging for us to evaluate as we are not conducting a temporal analysis of the networks. However, based on the previous metrics mentioned, we will be able to identify different types of users based on their participation and identify the current percentage of active users across different channels. The node level metrics can help use identify the different categories of users we have across the networks. Since PICUP has been established for a fair number of years, we would expect the networks to have a fair amount of users that are highly active to indicate there are leaders beyond the creators of PICUP, at least in the Slack community. Additionally, if PICUP was increasing the number of community leaders, we would expect many active broadcasters in the network who are sharing and facilitating information with other users. 

In addition to the metric values, we can identify these features in the network representations. Ideally, we would see many users with a large node size and light blue color, indicating their active status in the network. Between these users, there would be grey edges with large weights to emphasize frequent, reciprocal communication. Opposingly, we would see few users with small strength and small in strength on the periphery of the network. Ideally, we would expect a mix of messages sent to the entire channel and messages sent directly to another user. This balance would highlight a cohesive community in the Slack channel with users that feel comfortable chatting with each other as well as the expert knowledge of specific users in the community. 

In Section~\ref{sec:SNA}, we will assess whether PICUP has met the three identified goals on three networks using the metrics defined in Section~\ref{sec:SNAmetrics}. We conduct our analysis on three networks of varying size - both in number of users and messages - and varying channel topic focuses from pedagogy to computing. We will extend this analysis to five more channels to characterize the Workspace as a whole. 


As we will discuss more in the Section~\ref{sec:limitations}, we are limited by the claims we can make on whether these goals have adequately been met as we are only looking at the Slack community, which is one aspect of the PICUP community. There are, of course, other types of analyses we can use to further assess PICUP's goals.

\section{Assessing PICUP Goals Using Social Network Analysis}
\label{sec:SNA}
Before beginning our analysis, we note that the users of PICUP can join the Slack channel at anytime and thus, the network is a snapshot of the communication in the channel up to that point. This framing influences how we analyze the networks. While a network may contain less active users on the periphery, this does not necessarily indicate those users will always remain less active. They simply joined the channel shortly before we ended our data collection. 

In the following sections, we analyze eight channels with a variety of topic focuses, number of users, and number of messages. We chose three of the eight to perform a deeper analysis on - Advanced Thermodynamics, Jupyter, and Classroom Pedagogy. As we will explain in each section, the three channels represent three different network sizes (in terms of number of users and number of messages) and have three distinct discussion focuses.  

\subsection{Advanced Thermodynamics}\label{sec:AT}
The Advanced Thermodynamics channel is nearly the smallest channel in the Slack Workspace with 16 users, 60 raw Slack messages, and 678 messages post data restructuring. While the amount of nodes and edges in this network is on the smaller side, we can still retrieve meaningful information from the analysis. The channel is pedagogically focused with users discussing resources, ideas, and strategies for integrating computing into their Advanced Thermodynamics curriculum. While users are talking about computing, the emphasis is on the integration of computing into the task rather than the computing tool directly. As we will see in each of the networks described, there are three clear subcategories representing three different levels of engagement in the network.

\begin{figure*}[htp]
    \centering
    \begin{subfigure}[b]{0.23\textwidth}
        \centering
        \includegraphics[width=\textwidth]{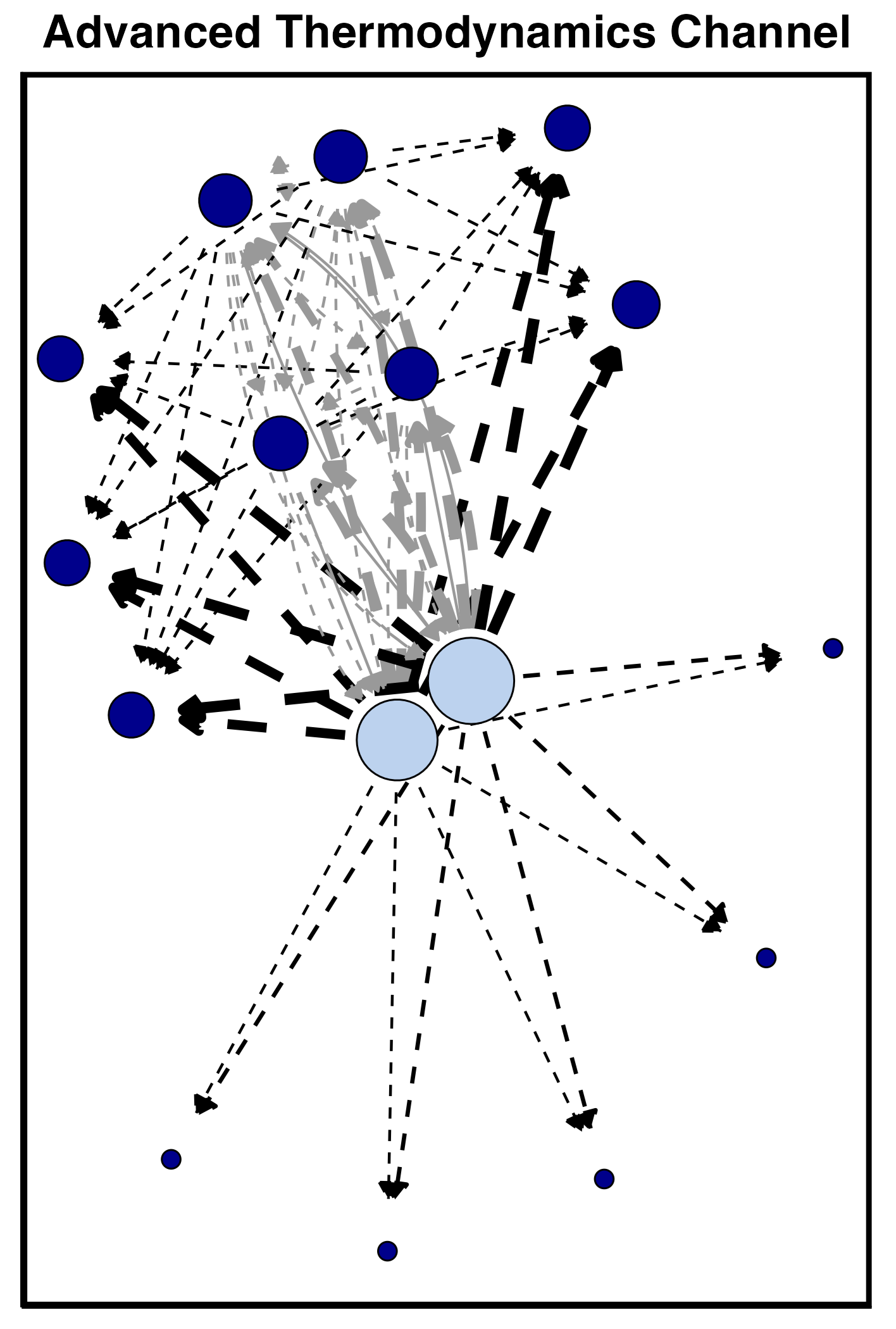}
        \caption{}
        \label{fig:dwc_advtherm}
        \end{subfigure}
    \begin{subfigure}[b]{0.23\textwidth}
       \centering
       \includegraphics[width=\textwidth]{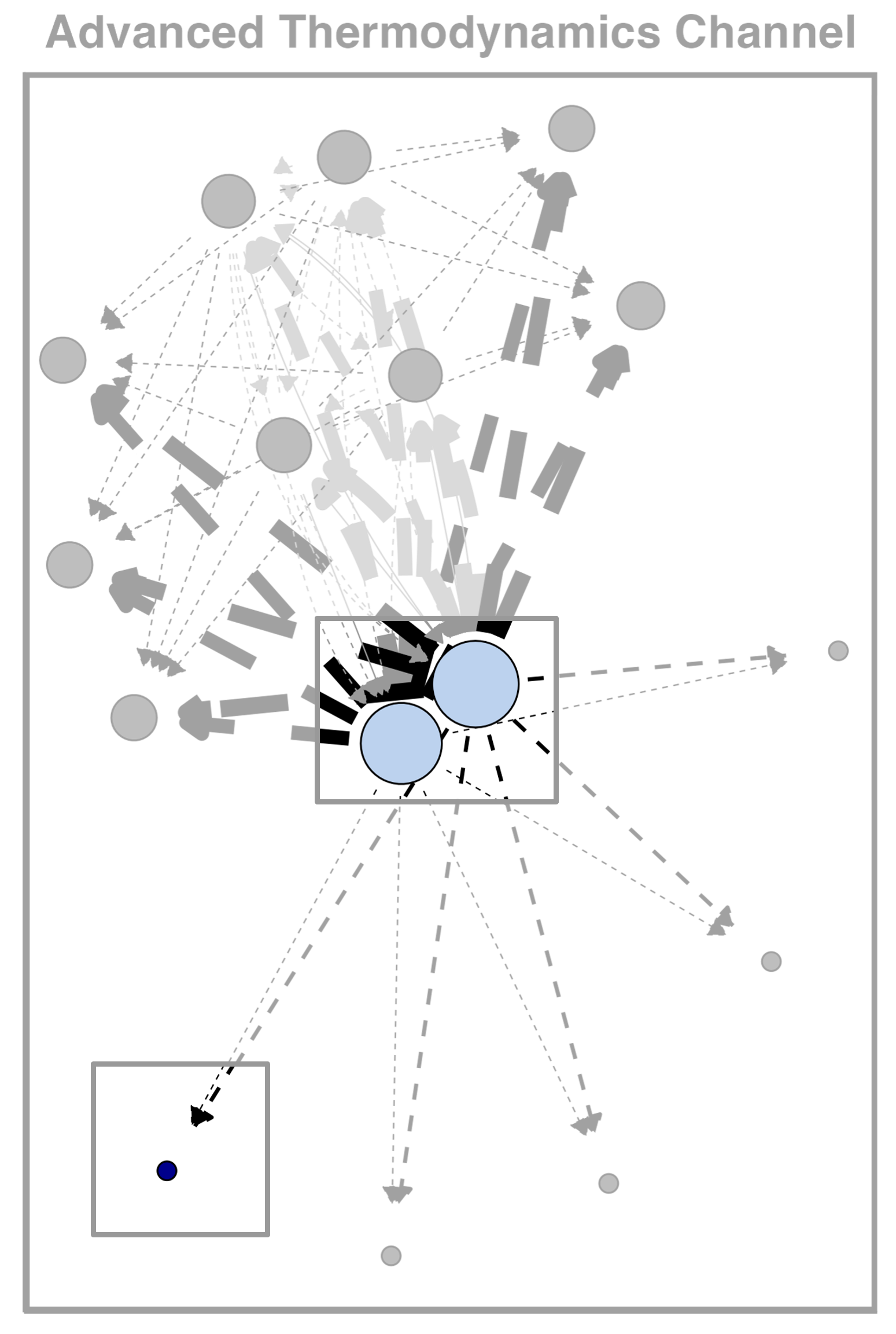}
       \caption{}
       \label{fig:pt1} 
    \end{subfigure}
    \begin{subfigure}[b]{0.23\textwidth}
       \centering
       \includegraphics[width=\textwidth]{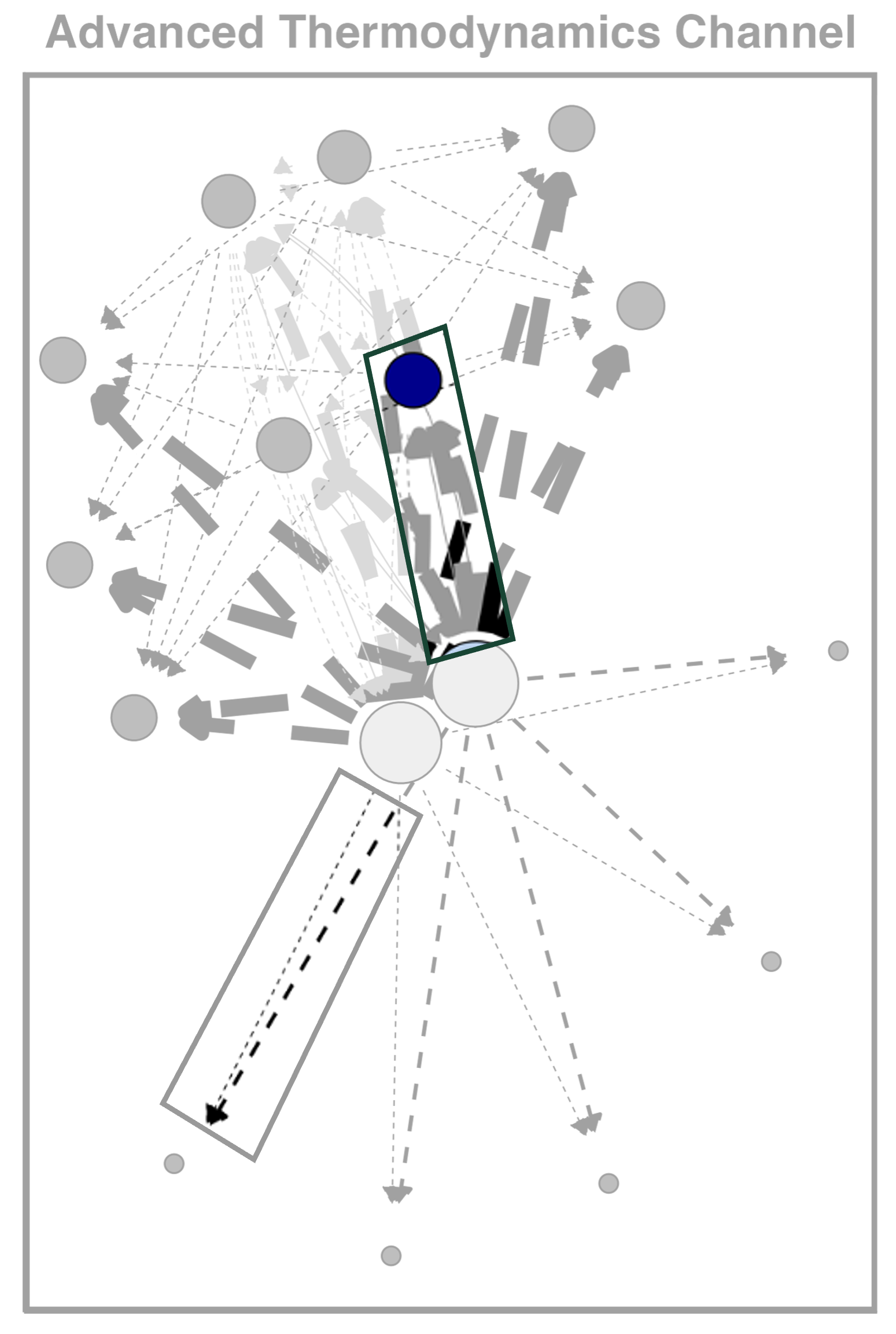}
       \caption{}
       \label{fig:pt3}
    \end{subfigure}
    \begin{subfigure}[b]{0.23\textwidth}
       \centering
       \includegraphics[scale=0.42]{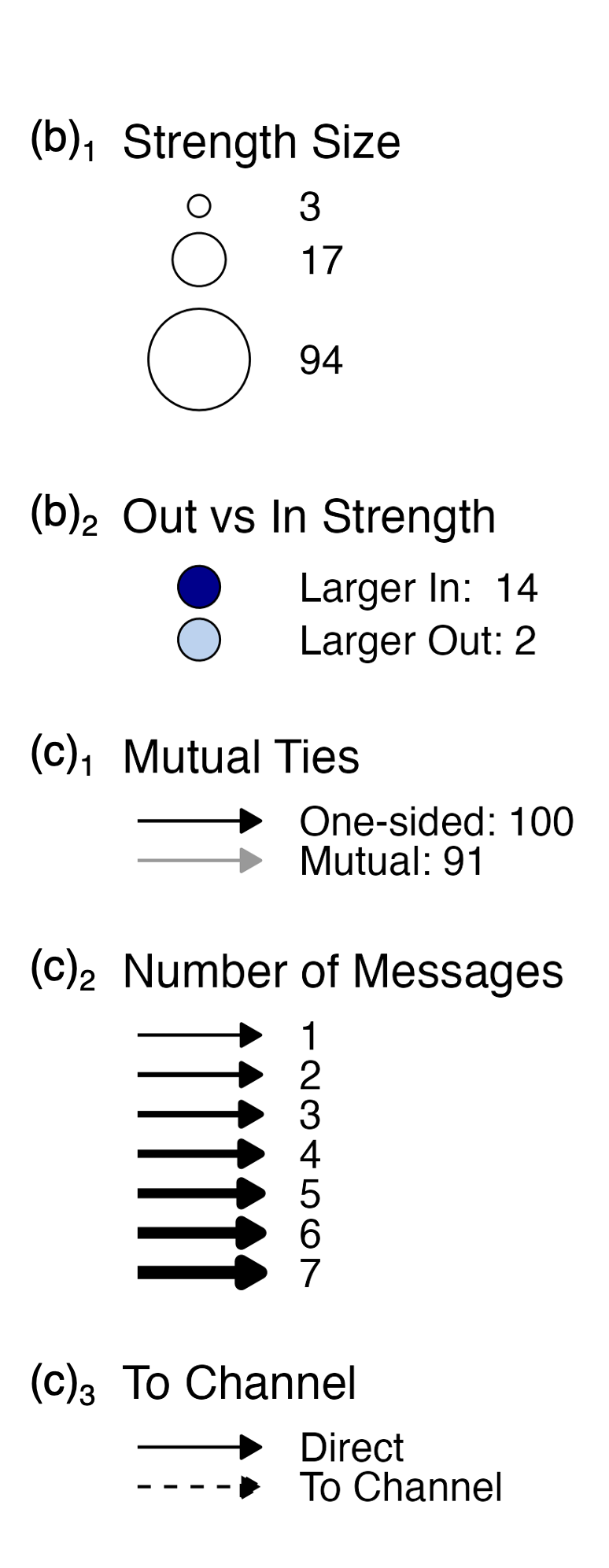}
       \caption{}
       \label{fig:legend}
    \end{subfigure}
       \caption{Figure~\ref{fig:dwc_advtherm} shows the weighted, directed network for the Advanced Thermodynamics channel. Using the legend in Figure~\ref{fig:legend}, we highlight the representations for nodes (Figure~\ref{fig:pt1}) and edges (Figure~\ref{fig:pt3}) in the network. The three representations show the same network.}
       \label{fig:all_tog}
 \end{figure*}

Extending the representation from Figure~\ref{fig:example_network}, we represent the Advanced Thermodynamics channel with a directed, weighted network in Figure~\ref{fig:dwc_advtherm}. We use the rest of Figure~\ref{fig:all_tog} to connect the elements of the Figure~\ref{fig:example_network} to the network representations.  Figure~\ref{fig:pt1} and legend objects $(b)_1$ and $(b)_2$ depict node size and color. As mentioned earlier, larger node sizes represent a larger strength value and nodes with a larger out-strength than in-strength are light-blue in color. Figure~\ref{fig:pt3} and legend objects $(c)_3$, $(c)_4$, and $(c)_5$ represent non-reciprocal (black edges) and reciprocal (grey edges) communication, show edge weight (the number of messages between the set of nodes), and depict whether a message is sent to the entire channel (dashed edge) or to a user directly (solid edge). While this representation breeds complexity and additional explaining, it also allows for richer contextualization of the relationships in the Slack channel. We note all networks use the Kamada-Kawai (KK) layout in igraph \citep{kk}. 

In the Advanced Thermodynamics channel, there are two light blue nodes. These highly active users in the network have the largest strength, the largest out-strength, and the largest CC\_{out} ($32.2$, $27.7$), and a large CC\_{in} ($5.18$, $5.09$). As expected, there are many edges - both grey and black - connected to these nodes. With their large strength and large closeness centrality values, these users - who are in PICUP leadership - are in frequent communication with other users and are broadcasters of information.  

The two highly active users are connected to four other users through many grey ties. These users have a larger in-strength than out-strength and have some of the largest CC\_{in} values between $5.8$ and $6.6$, indicating they are frequent receivers of information. The six users connected by reciprocated edges signify a sub-community within the network. 

Extending this part of the network to the five dark blue nodes with similar strength, we identify a third collection of users that have an out-strength of $0$. These users have the highest CC\_{in} values of $6.85$. The users on the lower right of the network also fall into this category, they have joined the channel later and thus, they have a smaller strength. While they have large CC\_{in} values and have the similar knowledge to other users in the channel, they are more passive members of the community.  

Looking at the legend, despite only 2 users sending more messages than they are receiving, we have about the same amount of one-sided (100) and mutual messages (91). This is due to the sheer number of messages sent to the whole channel with only about 6 users engaging in reciprocal conversations. This could be indicative of the users not using the tagging feature correctly, or truly wanting to share a message with the entire channel.

\begin{figure*}[htp]
    \centering
    \begin{subfigure}[b]{0.45\textwidth}
        \centering
        \includegraphics[width=\textwidth]{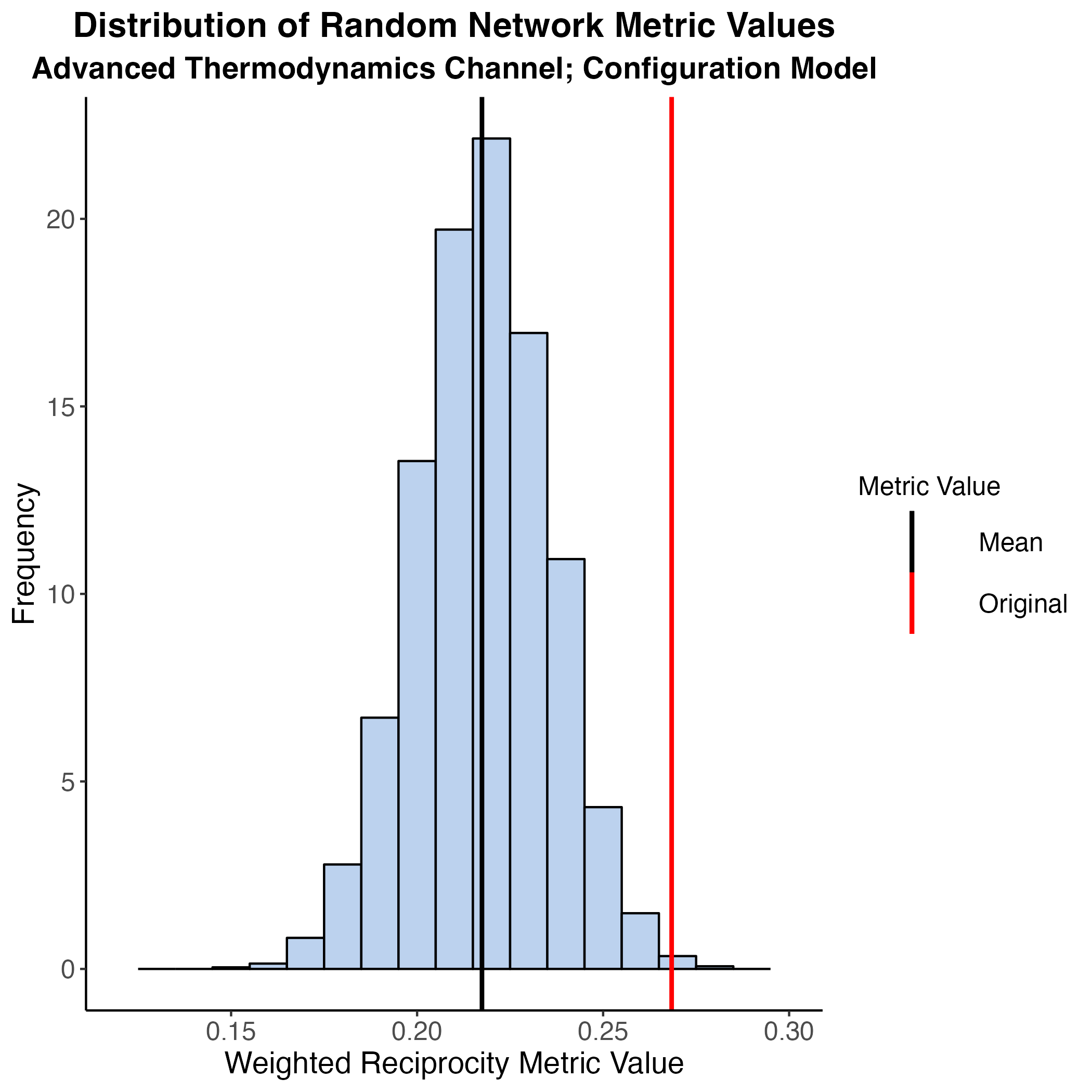}
        \caption{ }
        \label{fig:AT_config_wr_DistRandom}
    \end{subfigure}
    \hfill
    \begin{subfigure}[b]{0.45\textwidth}
        \centering
        \includegraphics[width=\textwidth]{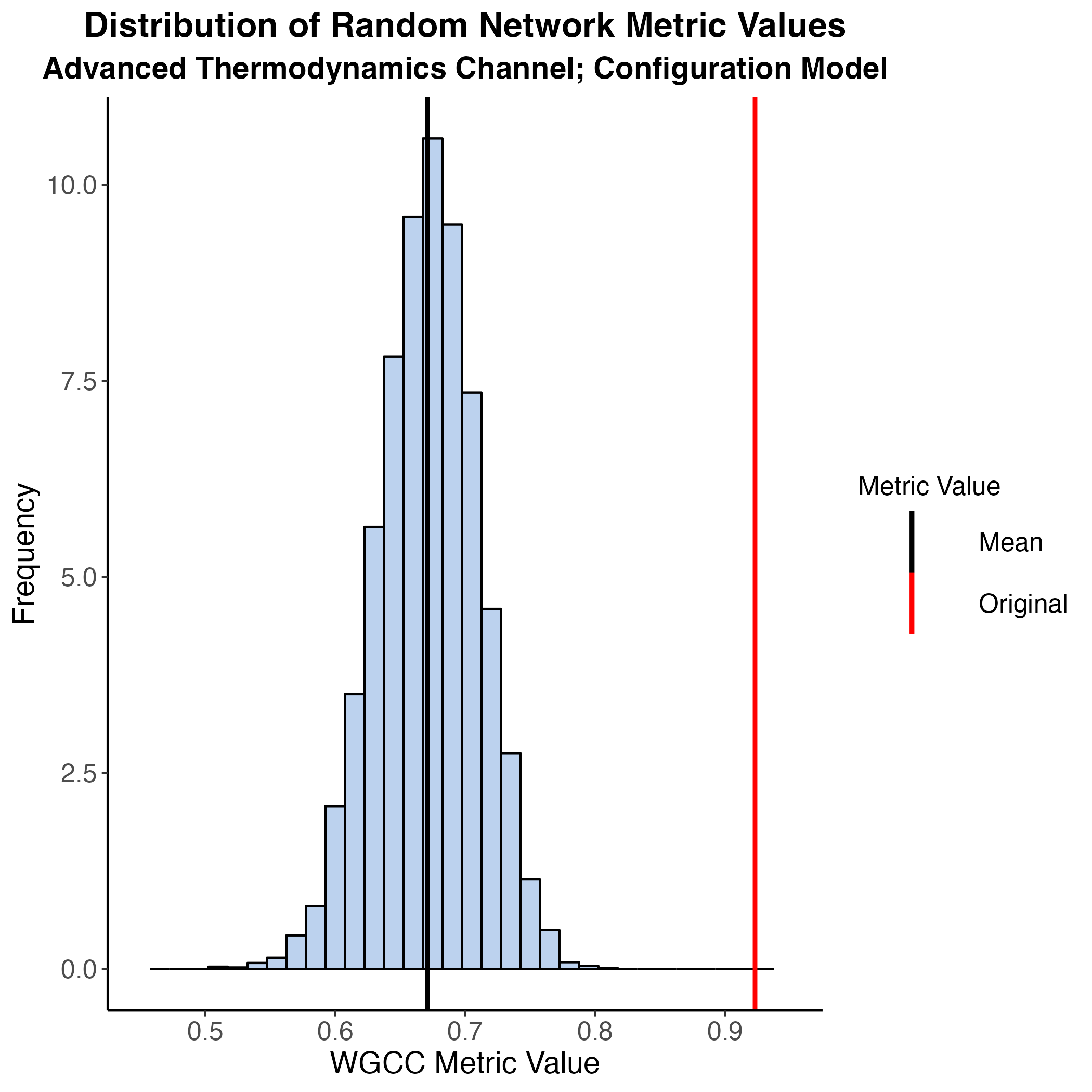}
        \caption{  }        
        \label{fig:AT_config_wgcc_DistRandom}
    \end{subfigure}
    \caption{Applying the Configuration Model to Advanced Thermodynamics Network with the weighted reciprocity and weighted global clustering coefficient metrics. The probability value for both is large indicating our original network has a much larger weighted reciprocity value and weighted global clustering coefficient compared to networks of similar size and distribution. } 
    \label{fig:AT_DistRandom_Config}
\end{figure*}

The weighted reciprocity value for the Advanced Thermodynamics network is $0.268$. Looking at Figure~\ref{fig:AT_config_wr_DistRandom}, this value is higher than the average value of the ensemble with a probability value of $0.996$. This implies the reciprocity in our network is higher than other, similar networks in the ensemble. However, with the weighted reciprocity metric bounded between $0$ and $1$, our value of $0.268$ is on the lower end. Looking closely at the reciprocal ties in the network, there are not many reciprocal relationships, and the reciprocal relationship that do appear are incredibly one-sided. In fact, most of the reciprocal ties between two users are highly imbalanced - an edge with a large weight for one user and an edge with a small weight for the reciprocal tie. Recall, the weighted reciprocity matrix is calculated by creating a symmetric network with minimum weight between the edges for two users, thus, lowering the weighted reciprocity value. 

From Figure~\ref{fig:AT_config_wgcc_DistRandom}, the WGCC is $0.923$. This value indicates the edges in the network that appear in closed triplets have a higher edge weight than those that do not. Further, nearly all the triplets present in the network are closed (like (3) in Figure~\ref{fig:triplets_dir}). This is fairly easy to see in the network. Most of the interactions from the middle to the top left are a part of at least one closed triplet. However, the users along the periphery in the bottom right are not connected to each other and likely represent the open triplets pushing the weighted GCC from $1$. The high closed triplet value indicates intercommunication between clusters of three people in the network. However, most of the messages were sent to the entire channel. This makes it challenging to identify true clusters of users in the network, rather than those who just are a part of triplets due to sending messages to the entire channel. With this nature of messaging, the likelihood that a user is in a closed triplet is correlated to the time they joined the channel. This type of high value for clustering is typical in social networks \citep{newmanmej2003whysocialnetworksaredifferent}. 

Looking at the distribution of WGCC values for the ensemble created with the configuration model, the average WGCC is about $0.675$ which is much smaller than the true value. With a probability value of $1$, the WGCC for the Advanced Thermodynamics network is larger than all other networks in the ensemble.

\subsection{Jupyter}\label{sec:JT}

\begin{figure}[htp]
    \centering
    \includegraphics[width=0.45\textwidth]{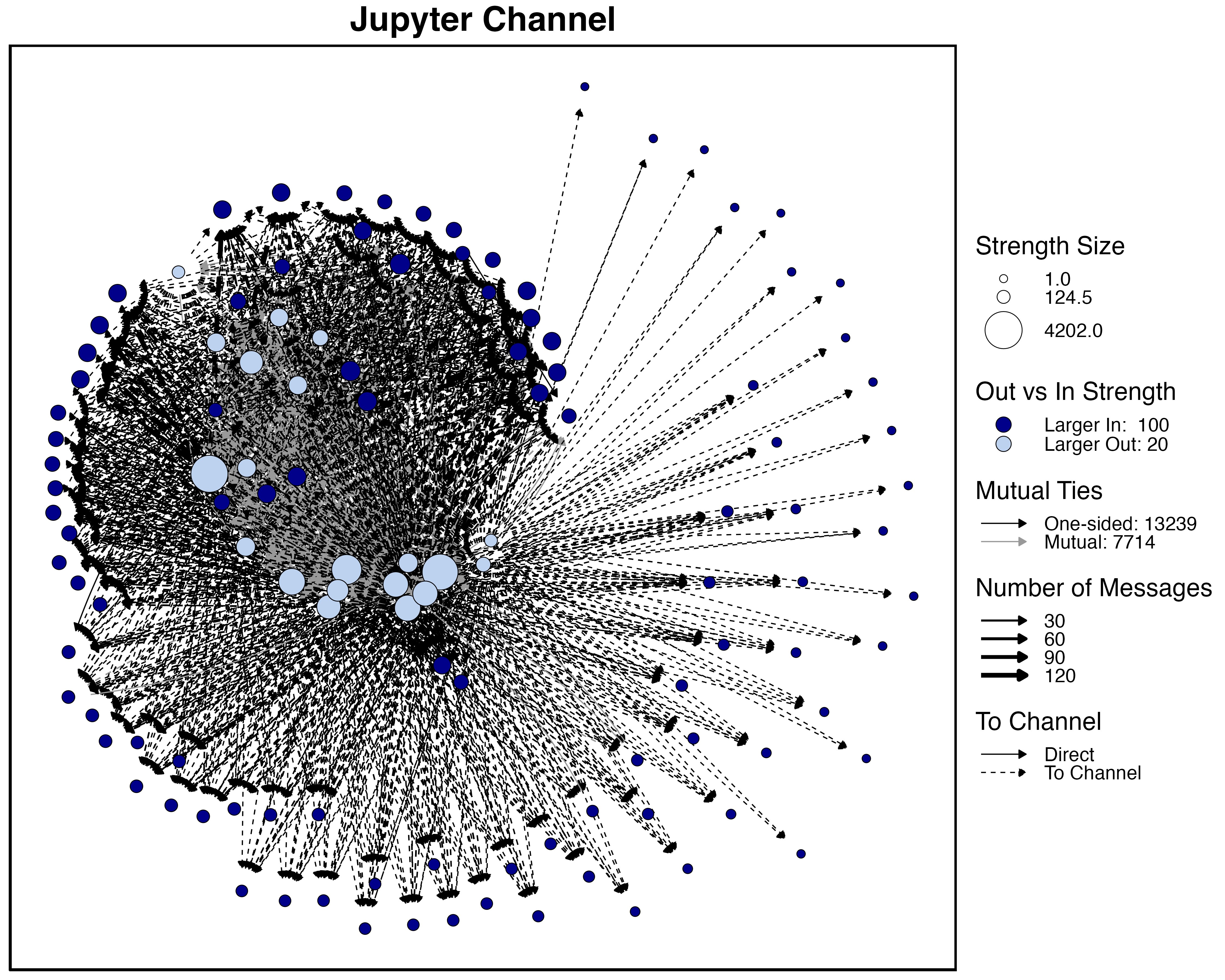}
    \caption{We show the directed, weighted network representation of the Jupyter Channel. Despite the drastic increase in size from the Advanced Thermodynamics network, we see similar network characteristics with three levels of users engagement.}
    \label{fig:dwc_jupyter}
\end{figure}

Besides `General' and `Random', which by default include everyone in the Slack Workspace, `Jupyter' is the largest channel in the workspace with 119 users, 777 Slack messages, and 28,926 edges in the post processed data. The channel is used to discuss questions and ideas related to Jupyter - an open-source software used for interactive computing with coding languages like Python, Julia, and R. Most commonly, it is used in Jupyter Notebooks. The interface intertwines coding blocks with written explanations in one file. With a low barrier to entry, they are a great pedagogical tool. 

The Jupyter channel is a contrast to the Advanced Thermodynamics channel. Jupyter is one of the largest networks in the channel with 4x as many users and 43x as many edges as Advanced Thermodynamics. Additionally, the channel discussion is focused on a specific software tool rather than a specific course or field. While there is likely overlap in these channels are far as how to integrate this software into coursework, this channel is more heavily focused on usage of the software including troubleshooting and error handling. 

The weighted and directed network for the Jupyter channel can be seen in Figure~\ref{fig:dwc_jupyter}. While it is challenging to see minute details with the large number of edges, we can still extract overall structure from the network. About 16\% of the users have a larger out-strength, with about 35\% of those users are within PICUP leadership. These users also have extremely large CC\_{out} values. We notice few reciprocal edges in the top left portion of the network between the highly active users. Opposingly, 83\% of the users on the periphery of the network in the channel are receiving more messages than they are sending. A sub group of these users have a non-zero out-strength and have some reciprocated communication with other users in the network, but nearly 50\% of the users have an out-strength of $0$. Further, there are 1.7x more non-reciprocated edges than reciprocated edges - many of which where sent to the entire channel.  

In general, we are seeing a similar structure to the Advanced Thermodynamics channel. There are three types of users, with varying levels of engagement, and most of the communication in the network is passive and one-sided as there are many black edges.

\begin{figure*}[htp]
    \centering
    \begin{subfigure}[b]{0.45\textwidth}
        \centering
        \includegraphics[width=\textwidth]{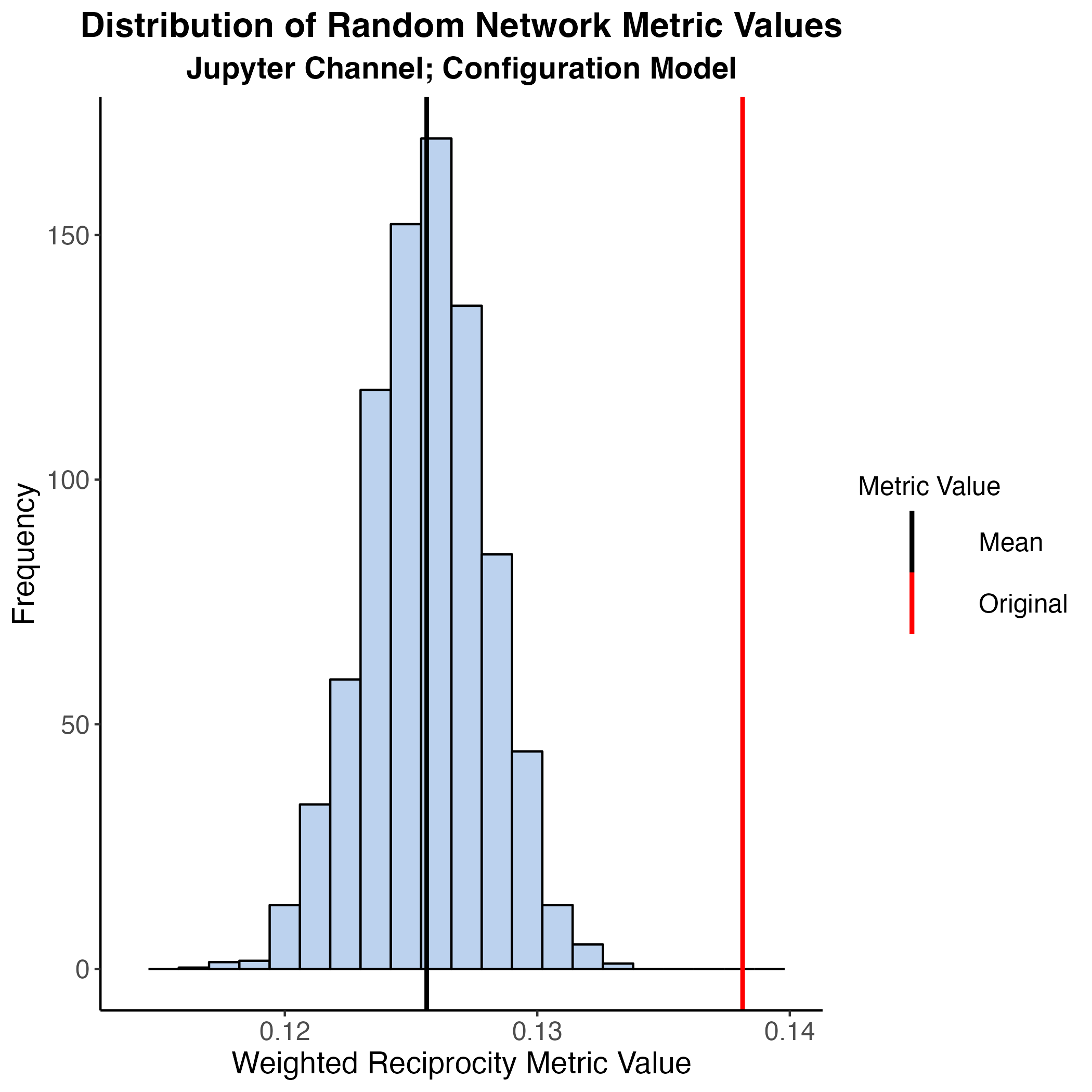}
        \caption{}
        \label{fig:J_config_wr_DistRandom}
    \end{subfigure}
    \hfill
    \begin{subfigure}[b]{0.45\textwidth}
        \centering
        \includegraphics[width=\textwidth]{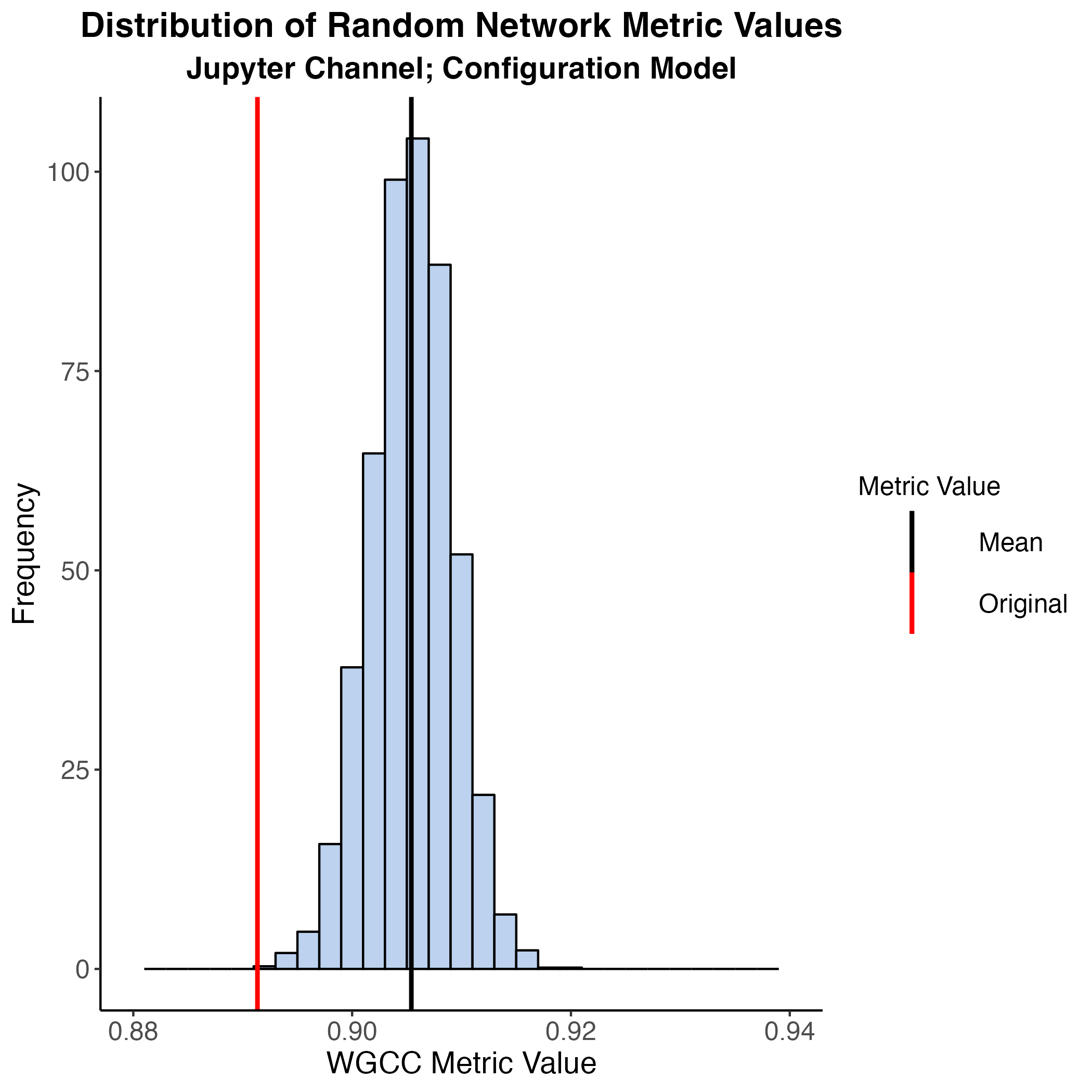}
        \caption{  }        
        \label{fig:J_config_wGCC_DistRandom}
    \end{subfigure}
    \caption{We apply the Configuration Model to the Jupyter network, calculating the weighted reciprocity and WGCC metrics. For weighted reciprocity, the probability value is $1$ indicating more reciprocal communication than in the ensemble. For WGCC, the probability value is $0$ indicating the value is smaller than other networks in the ensemble. } 
    \label{fig:J_DistRandom_Config}
\end{figure*}

For the whole network metrics, we see similar values to the Advanced Thermodynamics networks. Looking at Figure~\ref{fig:J_config_wr_DistRandom}, the weighted reciprocity of $0.138$ is higher than all other networks in the ensemble with probability $1$ at a value. While the value is large comparatively, it is low highlighting the lack of reciprocity network, particularly with the edges for the many nodes on the periphery. Additionally, the WGCC is $0.891$ which is relatively high, but comparatively speaking lower than the networks in the ensemble. 



\subsection{Classroom Pedagogy}\label{sec:CP}
Finally, Classroom Pedagogy is a medium-sized channel with 72 users, 102 raw Slack messages, and 3,338 messages post data restructuring. The channel provides a balance in the number of users and messages between the previous two channels, and also is focused solely on pedagogical practices. While channels like Trinket and Glowscript may be a better network size comparison between Advanced Thermodynamics and Jupyter, their channel topics are too closely aligned with Jupyter. 

\begin{figure}[htp]
    \centering
       \includegraphics[width=0.45\textwidth]{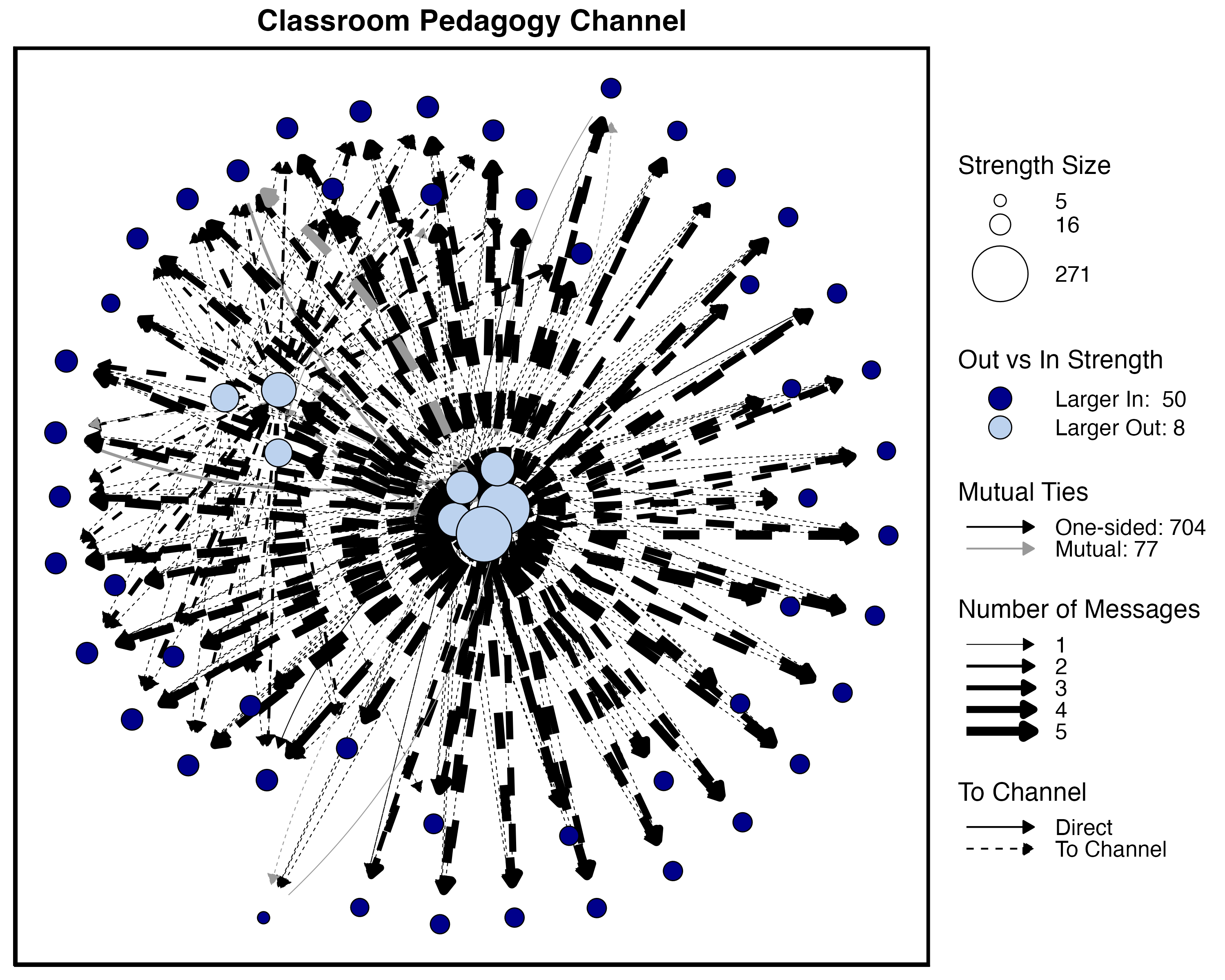}
       \caption{ The Classroom Pedagogy channel as a  network. We see similar trends to the other networks in this paper. There are few highly active users and many users on the periphery.  }
       \label{fig:CP} 
 \end{figure} 

In the Classroom Pedagogy network, Figure~\ref{fig:CP}, we see similar structure to the previous two networks in relation to the types of users. There are a small portion of users that are highly active with large out-strength and large CC\_{out} values, a larger portion of users with a larger in-strength than their non-zero out-strength, and an even larger portion of users with a $0$ out-strength. Despite the similar categorization of users, the proportion of reciprocated edges to non-reciprocated edges is much smaller - about 10\%. This is also visible through the network which has a highly radial structure with many black edges. 

\begin{figure*}[htp]
    \centering
    \begin{subfigure}[b]{0.45\textwidth}
        \centering
        \includegraphics[width=\textwidth]{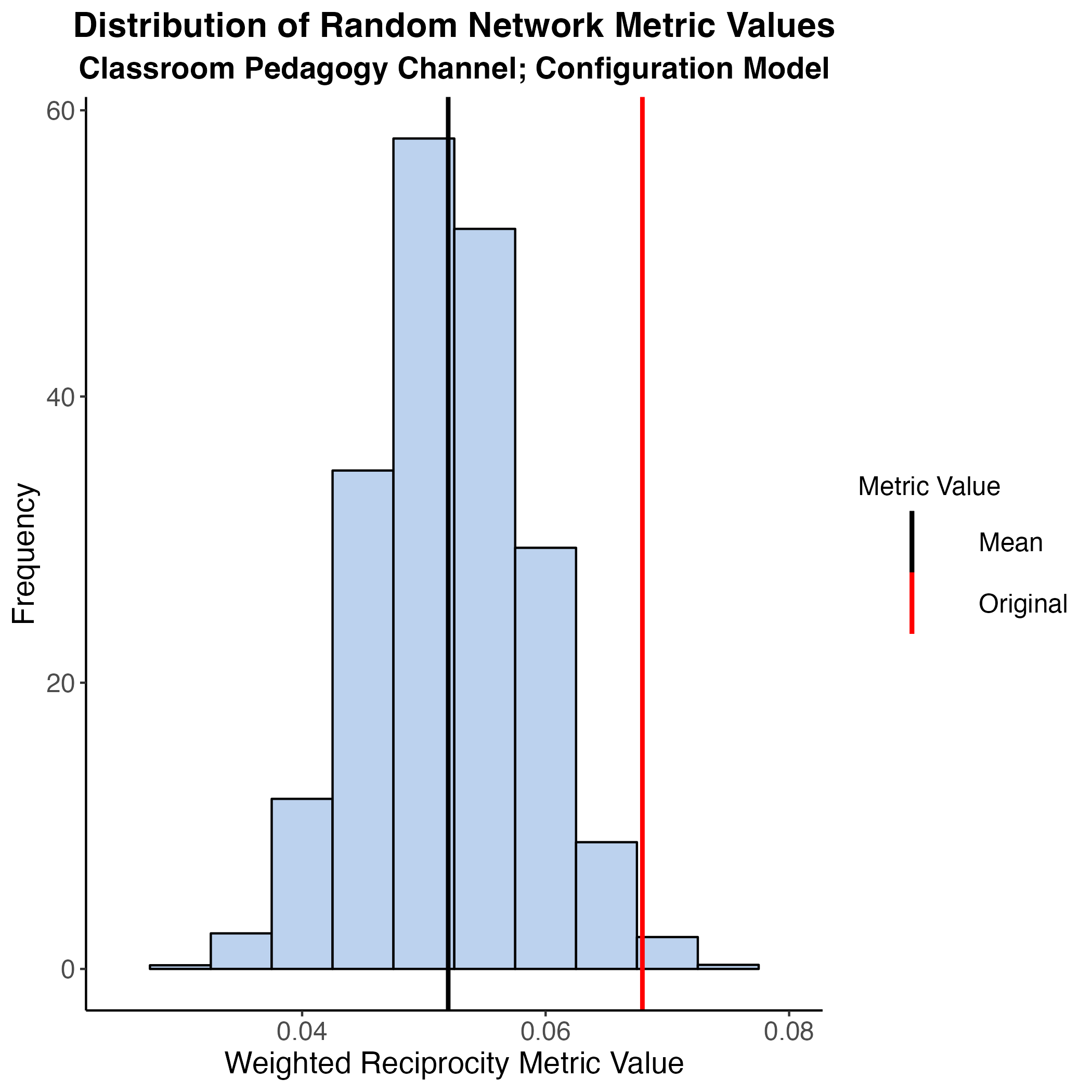}
        \caption{}
        \label{fig:CP_config_wr_DistRandom}
    \end{subfigure}
    \hfill
    \begin{subfigure}[b]{0.45\textwidth}
        \centering
        \includegraphics[width=\textwidth]{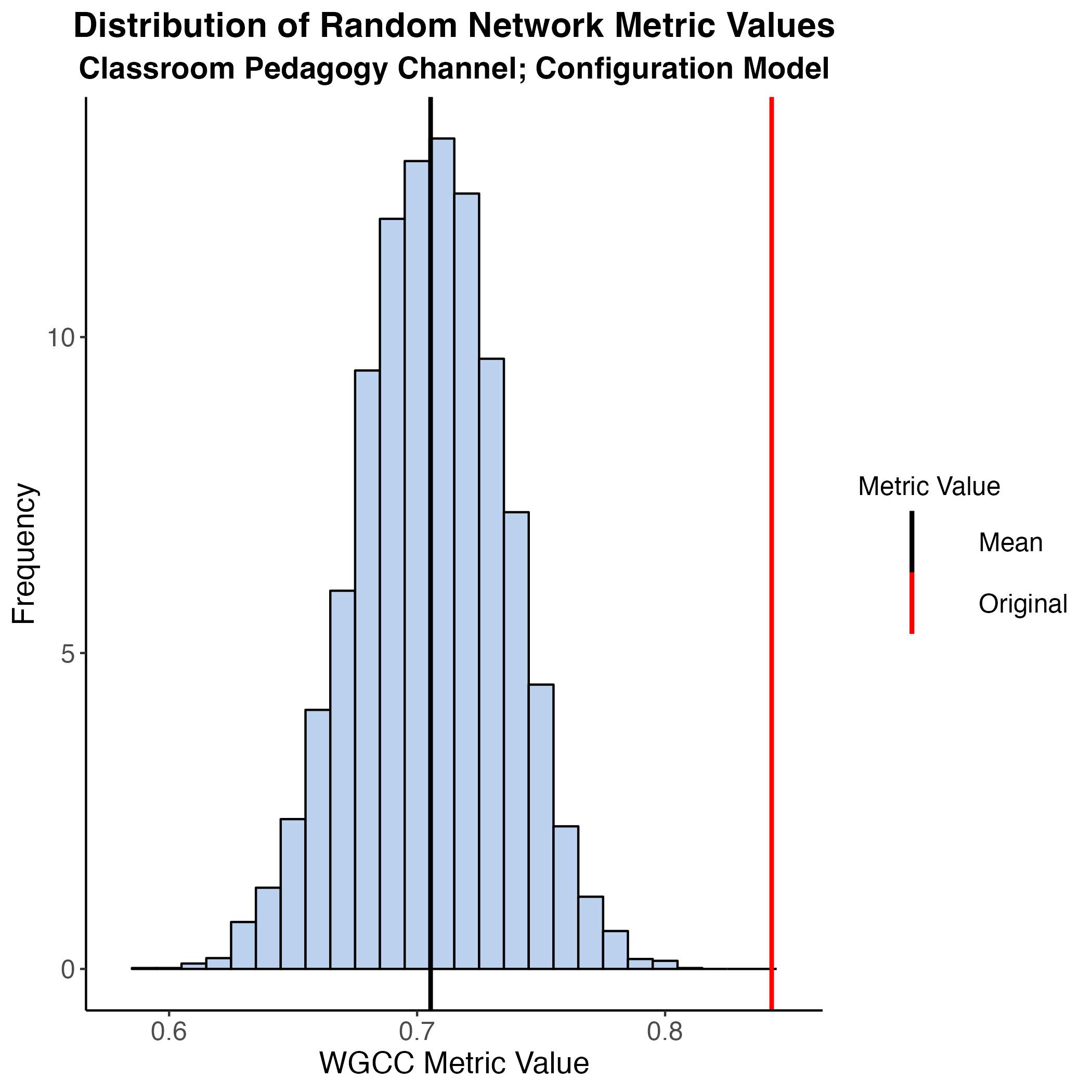}
        \caption{  }        
        \label{fig:CP_config_wGCC_DistRandom}
    \end{subfigure}
    \caption{We apply the Configuration Model to Classroom Pedagogy channel. For Weighted Reciprocity, the probability value is $0.98$ and for WGCC, the probability value is $1$. Both values indicate the metric values are larger than other, similar networks in the ensemble.  } 
    \label{fig:CP_DistRandom_Config}
\end{figure*}

In line with the previous two networks, the weighted reciprocity and the WGCC for the Classroom Pedagogy channel are $0.068$ and $0.843$ respectively, as seen in Figure~\ref{fig:CP_DistRandom_Config}. With probability $1$, both are greater than similar networks in the ensemble generated by the Configuration Model. Again, the weighted reciprocity value is low, emphasizing the lack of reciprocal communication as seen in the network. The WGCC value is lower than the previous two networks, but still relatively high indicating a fair amount of closed triplets in the network.

\subsection{Other Channels}\label{sec:other}

In each of the three networks in Sections~\ref{sec:AT}-\ref{sec:CP}, we saw three different types of engagement:
\begin{enumerate}
    \item Active Users - Nodes with a larger out-strength than in-strength 
    \item Passive Users - Nodes with a larger in-strength than out-strength, and a nonnegative out-strength
    \item Receivers - Nodes with nonzero in-strength, and an out-strength of 0. 
\end{enumerate} 

Using these categories, we calculated the percentages of users in each group across eight channels in the Slack Workspace - seen in Table~\ref{tab:all_channels}. While not an in-depth analysis, we can explore trends of participation across multiple channels in the PICUP Slack. The Receivers category is consistently the largest percentage of users (or equivalent to Passive Users in the Upper MidWest channel). Often, the percentage of users in the first group is smaller than the second, with Advanced Thermodynamics, Trinket, and Jupyter as the exception. However, even in those channels the average difference between the first and second categories is 8.3\%.

\begin{table*}[htp]
    \centering
    \caption{Percentages for users in each of the three defined categories across eight channels in the PICUP Slack Workspace. The third category nearly always has the largest percentage of users, and the first category is the smallest in all but three channels.}
    \label{tab:all_channels}
    \begin{tabularx}{\linewidth}{m{3cm}||m{1.75cm}|m{1.75cm}|m{1.75cm}|m{1.75cm}|m{1.75cm}|m{1.75cm}|m{1.75cm}|m{1.75cm} }\hline\hline
        Channel \newline Name & Advanced \newline Mechanics & Advanced \newline Therm & Upper \newline MidWest &  Why We \newline Do This &  Classroom \newline Pedagogy & Trinket & Glowscript & Jupyter \\ \hline
        Active Users & 16.7\% & 12.5 & 26.7 & 17.4 & 12.5 & 24.3 & 16.5 & 16.7\\ 
        Passive Users & 4.2 & 31.3 & 36.7 & 26.1 & 31.3 & 13.5 & 23.1 & 15.0\\ 
        Receivers & 79.2 & 56.3 & 36.7 & 56.5 & 56.3 & 62.2 & 60.4 & 68.3\\\hline\hline
    \end{tabularx}
\end{table*}

Looking at these categories on a user level, we calculated the number of users in the same category across multiple channels. We found that only $2$ users appear in the first category of more than $50\%$ of the eight channels and $20$ apppear in more than $25\%$ of the eight channels. Similarly, $10$ users appear in the third category of more than $50\%$ of the eight channels and $69$ apppear in more than $25\%$ of the eight channels. While there is more consistency in the users that remain in the third category, there are not many users that are using multiple channels. Additionally, only $12$ of the $174$ (or ~$7\%$) total users in the eight channels are in more than $50\%$ of the eight channels, with $8$ users only in $62.5\%$ of them. 


\section{Implications}
\label{sec:implications}

Extending from Section~\ref{sec:SNA}, we connect our findings from our Social Network Analysis to our Conceptual Framework to assess if PICUP is meeting their defined goals in the context of the Slack Workspace. 



The first PICUP goal we identified is: \textit{Lowering barriers for faculty to integrate computation into courses}. In the Slack, low barriers would indicate users are frequently interacting with one another to ask questions and share ideas. To align with this, we expect the network to have a fair amount of active users, particularly with high strength values. Across all networks we analyzed, the group of Active Users was often the smallest percentage of the three categories (or at least far smaller than the Receivers category). This corresponds to our visual assessment few light blue nodes. Often, these users seemed to be driving the communication in the channel with many of the reciprocal edges tied to them. This small group of users likely have a sense of community amongst themselves, but this clearly does not radiate through the rest of the channel. 

The second PICUP goal is: \textit{Continual community growth and strength in community building}. Similar to the first goal, we expect many active users in the network with high strength values. As mentioned, only a small percentage of users in each network are in the Active Users category. These few users have a high (out) strength and a high closeness centrality. The users send/receive many messages while also being good broadcasters of information to the rest of the channel. These users are essential to sharing information with the whole network and initiating conversations. While we do not expect all users to take on this role, having more users in this category would increase the amount of communication between users and strengthen the community. 

In addition, to better assess the community in the network, we expect high values for reciprocity and the global clustering coeffient to represent frequent communication across all users in the network. Across all channels in Sections~\ref{sec:AT}-\ref{sec:CP}, the weighted reciprocity value was larger than networks in the generated ensemble, but it was consistently low. 
With few Active Users and even few Passive Users, the low weighted reciprocity is not entirely surprising. As indicated by the few grey edges, there is little reciprociated communication in the networks, usually only between a small percentage of the users. The most active users in the network are often sharing messages with the community and receiving little to no responses. Opposingly, the weighted global clustering coefficient was fairly high for all three networks in Sections~\ref{sec:AT}-\ref{sec:CP}. While initially surprising given the other metrics, it is likely this metric is high due to the cluster of Active Users dominating the closed triplet total value as drivers of the conversation. 

The final PICUP goal we assess is: \textit{Increased growth in number of community leaders}. The PICUP community has now existed for many years, and while the core creators continue to be involved, the community is meant to expand by creating more local leaders which will further increase PICUP's reach. The mechanism for creating these leaders is often through in-person workshops, however, because the Slack community is an important aspect to PICUP, we would expect some of this leadership to translate to this domain. While we are not performing a temporal analysis, we would still expect to see these leadership represented through Active Users and even Passive Users, given the longevity of PICUP. Similar to the other two goals, we expect users to have high strength and high closeness centrality values. As explained, there are few users in each network that fit this description. 

Combining these findings, we argue PICUP is currently not meeting the three defined goals in the context of the Slack Workspace. For our three representative networks in Sections~\ref{sec:AT}-\ref{sec:CP}, we consistently observed low reciprocity, few active users, few broadcasters, many receivers, and many messages sent to the whole channel. In conjunction with our percentage calculations, the majority of users were Receivers in all eight categories. This domination of users not interacting with other users in the network through messages or reactions highlights the lack of whole network community and the number of community leaders. Further, it does not appear the PICUP community is engaging with one type of channel more than the other. This lack of interaction is consistent across channels focused on coding software \textit{and} specific curriculum. This extends across multiple networks as well. As we found, only 7\% of the total users in the Slack are in more than 50\% of the channels. Users are self-selecting which channels they would like to interact with and then choosing not to engage through messages or reactions in those channels. 

Together, these findings indicate a lack of community between the members of PICUP in the Slack Workspace. With few active users and many receivers, members of PICUP are not engaging in conversation with other members. However, this analysis does not lend itself to understand why this may be the case and what intentions PICUP creators and users have for the Slack.

\section{ Limitations and Future Work}
\label{sec:limitations}

As hinted at so far, while PICUP is not yet meeting their goals through our use of SNA to analyze the Slack community, there are limitations with this work and making broader claims. In this work, we only looked at the PICUP community interactions through Slack. As mentioned in Section~\ref{sec:introduction}, the PICUP community has many mechanisms for interaction through both in-person and virtual means, beyond the Slack community. By limiting the analysis solely to the Slack interactions, we are limiting the interpretations we can make with regards to PICUP meeting its goals. Moreover, the use of Slack itself might be a limitation. While it is a popular platform for online communities, it may not be the best fit for PICUP's goals. The nature of Slack encourages quick, informal communication, which may not foster the deep, meaningful interactions that PICUP aims for. This could explain the high number of messages sent to the entire channel without much engagement from other users.

PICUP is a complex and unique community. As described in the Section~\ref{sec:cf}, it does not fit squarely into traditional CoP and FOLC frameworks. The lack of rigid structure allows for more fluidity in engagement from users, but makes it more challenging to critically assess how each component of the community aligns with PICUP's broader goals. The flexible boundaries for users hinders the ability to define who is a part of the community, in what ways they interact with community, and what their goals are for being in the community. Additionally, while PICUP's broader goals are clearly defined, their specific goals for the Slack community are not. In this work, we constructed our own framework (in Section~\ref{sec:cf}) to use Social Network Analysis techniques to evaluate PICUP's overarching goals as it relates to the Slack Workspace. We found the engagement on Slack did not meeting PICUP's overall goals. However, PICUP's goals for the Slack community might be different or more nuanced than their overarching goals, which may alter our assessment. 

PICUP might consider better defining their intentions for the Slack Workspace and how it is integrated into their community. Through qualitative analyses, researchers could interview PICUP's creators and current leaders to understand their intended purpose for creating the online community. Similarly, researchers should include responses from community members on their needs and usage of Slack as they likely have different goals. Through clearer definitions, the PICUP community could more accurately evaluate whether they are meeting their goals and align their actions accordingly. It may be the case that Slack is simply meant to be a space for folks in the community to interact in an offline format. However, if PICUP is looking to more closely translate the in-person experience to the offline format, PICUP would need to make adjustments to how the Slack Workspace is managed to assist in creating a cohesive group and more community leaders. 

In a similar vein, to accurately assess whether PICUP is meeting their goals in aspects like knowledge building and growth in leadership, we need to expand the data and type of analysis used in this work. There is an opportunity for the PICUP community to use analyses such as qualitative and/or natural language processing techniques to look at the content of the messages \citep{mchugh2020uncovering, WangWang2022, COSTASILVA2022111397, martinez2003combining}. Conducting interview analyses with current PICUP membership to gain insight on how they interact and engage with the material and support that PICUP offers. These type of analyses in conjuction with the SNA work presented will give PICUP a comprehensive assessment of their community. Not seeing a trend in the Slack community does not imply that PICUP is not meeting those goals through other mechanisms.



Within the data specifically, messages with text and emoji reactions are treated equally in the dataset. The internet and social media age has altered how we communicate and share information with another. An emoji response represents an acknowledgement or reaction to a message, and thus, still represents a form of communication between those two users. While this type of engagement might be different than a traditional, written response, denoting this in the data would have created additional complexity in the network representations. Additionally, we chose to look at the the network at a snapshot in time rather than a time series progression, creating multiple networks at different points in time. This choice prohibited us from tracking the growth of the PICUP community and their interactions over time. It is worth noting that the time stamps in this type of analysis would needed to be on the order of months or years to see drastic progressions in the our data, especially for smaller channels. 

Further, we primarily chose to analyze each channel in silo. In this work, we compared the interactions of the PICUP community between different types of channels -- both in terms of size (both number of nodes and number of edges) and main topic for the network. We studied smaller, class-specific channels like Advanced Thermodynamics as well as larger, coding-focused channels like Jupyter. Interactions in one channel may influence relationships and interactions between specific users in other channels. To understand community interaction across networks, we calculated the percentage of users in each of the three engagement categories. This provided some insight into the participation of users throughout the channels, however, understanding the influence and interactions between specific users would require a much more in-depth node-level exploration \citep{Singh2025, wilson2009user}. Similarly, our inability to analyze private channels restricts our understanding of user engagement outside of public channels and how this may differ from their public engagement. 

In the networks, we chose to visually represent direct communication (solid) and to channel communication (dashed) separately, as they are distinct forms of communication. Multiple direct messages between two users may indicate a strong relationship between those two users, while multiple direct messages to a single user from many others in the channel might indicate an individual that has expertise in the given channel topic. While we cannot discern specifics about the interactions from the visual, representing the data in this way allows us to identify potential users this might be happening with. Similarly, messages sent to the whole channel both indicate there is a cohesiveness and strong comradery amongst the community, or indicate the community is fairly new with no identification of potential experts in the group. Again, it is not possible to conclude which is true based on the networks alone, but the distinction of the edge types starts these discussions. We note this difference in message content is not accounted for in the statistical analysis - all messages are treated the same in these calculations. 

Moreover, the weight of the edge accounts for the number of each of these type of messages. However, the metric values, such as strength, combine the total number of messages, and thus, do not take the type of message into account. While the edge type highlights a different type of communication, it is unclear if the users in the community are doing this properly - for example, are they appropriately tagging users they want answers from directly or are they messaging the whole channel without tagging a specific person? After looking briefly at the messages, it appeared the latter may be the case, and we decided to calculate the values for the different types of messages together. 

Further, we use grey edges to indicate reciprocal communication between the users. An edge is colored grey if there is a shared messaged between two users, either direct, to channel, or a combination of both. Qualitatively speaking, two private reciprocal messages may be different in meaning than two to channel reciprocal messages, but as mentioned, it is challenging to quantify this correctly if users are not properly tagging those they want to converse with. 

In addition to limitations and potential future analyses, we end this section with recommendations to increase engagement in the Slack Workspace. There are two main techniques for facilitating communication -- encouragement and guidance. Encouragement requires active users and leadership to encourage users to participate in activities related to the community \citep{tagarelli2018mining, preece2004top}. In PICUP, this could look like PICUP leadership encouraging users to share how they implemented a certain tool in the classroom or encouraging them to answer others' questions. This type of encouragement can build the confidence of users and hence, lead to increased use of the community. Through guidance, leadership and active users can help newcomers assimilate to the community quickly by setting norms, guidelines, and expectations for communication \citep{tagarelli2018mining, preece2004top}. As described by \citet{elliott2017evolution}, asynchronous communication is only as good as the users in it and as long as the interested is maintained. We note there is a lack of literature on strategies for encouraging participation in FOLCs in an equitable and inclusive way. General recommendations for online communitiy forums include inviting people from diverse communities and encouraging all users to introduce themselves \citep{abdullah2016equity}.

\section{ Conclusion}
\label{sec:conclusion}

In this work, we used Social Network Analysis to examine the communication between members of the PICUP Community on Slack and assess if their engagement aligns with PICUP's goals. We observed three levels of engagement for PICUP users: Active Users, Passive Users, and Receivers, where Receivers were the most prominent category across all networks analyzed. There were very few Active Users in each network with high strength and high CC values. These users dominated the conversation in each network through frequent reciprocal communication with other Active Users and broadcasting messages to the channel. However, across the whole network, there was generally low reciprocity values despite these values being larger than networks of similar size. 

The patterns observed indicate PICUP is not meeting its goals of Lowering Barriers, Community Growth, and Increased Community Leaders in the context of the Slack Workspace. There are few Active Users, particularly users beyond leadership, as well as limited reciprocal communication. Further, PICUP users do not frequently appear in multiple channels and tend to stay in the same engagement level across channels. Through the metrics and visually inspecting the networks, the Slack Workspace does not facilitate sustained community engagement in an online format. An ideal network with a cohesive community network would include more active users and reciprocal communication within and across networks. 

As mentioned in Section~\ref{sec:limitations}, the limitations of this research lend itself to future analysis. Our analysis of these goals is confined to Slack, and thus, future analyses such as text analysis and interview studies could complement this work and strengthen the assessment of PICUP's goals. Through a multi-method approach, researchers would understand the community's intentions and purpose for interacting with Slack and thus, PICUP could adjust their actions and guidance of engaging the community accordingly. 

We believe similar communities to PICUP can conduct an Social Network Analysis to assess the cohesion and engagement in their community. For other communities, we encourage clear definition of how the component being studied is situated in the context of larger community goals. Through our detailed study describing our choices for data preparation, network representation, conceptual framework, and  network calculation, we anticipate this work to serve as one example for other communities.

\section{Data Availability}
In compliance with Institutional Review Board, the deidentified data and code related to this project can be viewed in this \href{https://github.com/egbolger/PICUP_Publication_Codeshare.git}{Github Repository} \citep{egbolger}.

\section{Acknowledgements}
\label{sec:acknow}
Thank you to Daryl McPadden for their guidance on the early stages of this project. Thank you to Paul Irving and Savannah Wear for assistance with initial data collection and coding. This work was supported by: NSF DUE-1431776, NSF DUE-1504786, NSF DUE-1524128.

\bibliography{resubmittedbib.bib}

@inproceedings{silvia2019learner,
  title={A learner-centered approach to teaching computational modeling, data analysis, and programming},
  author={Silvia, Devin and O’Shea, Brian and Danielak, Brian},
  booktitle={Proceedings of the 19th International Conference on Computational Science–ICCS 2019, Faro, Portugal, 2019, Part V 19},
  pages={374--388},
  year={2019},
  organization={Springer, New York}, 
  doi = {10.1007/978-3-030-22750-0_30}
}

@article{magana2013introducing,
  title={Introducing discipline-based computing in undergraduate engineering education},
  author={Magana, Alejandra J and Falk, Michael L and Reese Jr, Michael J},
  journal={ACM Transactions on Computing Education (TOCE)},
  volume={13},
  number={1},
  year={2013},
  publisher={ACM New York, NY, USA}
}

@article{lee2020computational,
  title={Computational thinking from a disciplinary perspective: Integrating computational thinking in K-12 science, technology, engineering, and mathematics education},
  author={Lee, Irene and Grover, Shuchi and Martin, Fred and Pillai, Sarita and Malyn-Smith, Joyce},
  journal={Journal of Science Education and Technology},
  volume={29},
  pages={1--8},
  year={2020},
  publisher={Springer}
}

@article{chonacky2008integrating,
  title={Integrating computation into the undergraduate curriculum: A vision and guidelines for future developments},
  author={Chonacky, Norman and Winch, David},
  journal={American Journal of Physics},
  volume={76},
  number={4},
  pages={327--333},
  year={2008},
  publisher={AIP Publishing}
}

@article{caballero2019picup,
    author={Caballero, Marcos D and Chonacky, Norman and Engelhardt, Larry and Hilborn, Robert C and del Puerto, Marie Lopez and Roos, Kelly R},
    title={PICUP: A community of teachers integrating computation into undergraduate physics courses},
    journal={The Physics Teacher},
    volume={57},
    number={6},
    pages={397--399},
    year={2019},
    publisher={AIP Publishing},
    DOI={10.1119/1.5124281}
}

@article{leary2018difficulties,
  title={The difficulties associated with integrating computation into undergraduate physics},
  author={Leary, Ashleigh and Irving, Paul W and Caballero, Marcos D},
  journal={arXiv preprint arXiv:1807.03581},
  year={2018}
}

@article{henderson2019faculty,
  title={Faculty online learning communities: A model for sustained teaching transformation},
  author={Dancy, Melissa and Lau, Alexandra C and Rundquist, Andy and Henderson, Charles},
  journal={Physical Review Physics Education Research},
  volume={15},
  number={2},
  pages={020147},
  year={2019},
  publisher={APS}
}

@article{yadav2013professional,
  title={Who needs what: Recommendations for designing effective online professional development for computer science teachers},
  author={Qian, Yizhou and Hambrusch, Susanne and Yadav, Aman and Gretter, Sarah},
  journal={Journal of Research on Technology in Education},
  volume={50},
  number={2},
  pages={164--181},
  year={2018},
  publisher={Taylor \& Francis}
}

@article{smith2012predicting,
  title={Predicting continued use of online teacher professional development and the influence of social presence and sociability},
  author={Smith, Jo Ann and Sivo, Stephen A},
  journal={British Journal of Educational Technology},
  volume={43},
  number={6},
  pages={871--882},
  year={2012},
  publisher={Wiley Online Library}
}

@article{erickson2012effectiveness,
  title={Effectiveness of online professional development for rural special educators},
  author={Erickson, Amy S Gaumer and Noonan, Patricia M and McCall, Zachary},
  journal={Rural Special Education Quarterly},
  volume={31},
  number={1},
  pages={22--32},
  year={2012},
  publisher={SAGE Publications Sage CA: Los Angeles, CA}
}

@inproceedings{irving2017understanding,
  title={Understanding the PICUP community of practice},
  author={Irving, P and Caballero, M},
  booktitle={Physics Education Research Conference},
  year={2017}
}

@techreport{aapt,
author = ".", 
title = "AAPT Recommendations for Computational Physics in the Undergraduate Physics Curriculum", 
institution = "Project Report. The American Association of Physics Teachers, Undergraduate Curriculum Task Force", 
year = 2016
}

@techreport{phys21,
author = "Heron, Paula and McNeil, Laurie", 
title = "Phys21: Preparing Physics Students for 21st-Century Careers", 
institution = "Project Report. The American Physical Society and The American Association of Physics Teachers, Joint Task Force on Undergraduate Physics Programs", 
year = 2016
}

@book{wenger1999communities,
  title={Communities of practice: Learning, meaning, and identity},
  author={Wenger, Etienne},
  year={1999},
  publisher={Cambridge university press}
}

@article{tinnell2019sustaining,
  title={Sustaining pedagogical change via faculty learning community},
  author={Tinnell, Teresa L and Ralston, Patricia AS and Tretter, Thomas R and Mills, Mary E},
  journal={International Journal of STEM Education},
  volume={6},
  pages={1--16},
  year={2019},
  publisher={Springer}
}

@article{de2019developing,
  title={Developing and sustaining faculty-driven, curriculum-centered partnerships between two-year colleges and four-year institutions},
  author={De Leone, Charles J and Price, Edward and Sabella, Mel S and Van Duzor, Andrea G},
  journal={Journal of college science teaching},
  volume={48},
  number={6},
  pages={20--33},
  year={2019},
  publisher={JSTOR}
}

@article{furco2012using,
  title={Using learning communities to build faculty support for pedagogical innovation: A multi-campus study},
  author={Furco, Andrew and Moely, Barbara E},
  journal={The Journal of Higher Education},
  volume={83},
  number={1},
  pages={128--153},
  year={2012},
  publisher={Taylor \& Francis}
}

@article{price2021analyzing,
  title={Analyzing a faculty online learning community as a mechanism for supporting faculty implementation of a guided-inquiry curriculum},
  author={Price, Edward and Lau, Alexandra C and Goldberg, Fred and Turpen, Chandra and Smith, P Sean and Dancy, Melissa and Robinson, Steve},
  journal={International journal of STEM education},
  volume={8},
  pages={1--26},
  year={2021},
  publisher={Springer}
}

@book{cox2004building,
  title={Building faculty learning communities: new directions for teaching and learning, number 97},
  author={Cox, Milton D},
  volume={15}, 
  pages={5-23},
  year={2004},
  publisher={John Wiley \& Sons}
}

@article{baker1999creating,
  title={Creating learning communities: The unfinished agenda},
  author={Baker, Paul},
  journal={The social works of higher education},
  pages={10},
  year={1999},
  publisher={Thousand Oaks, California: Pine Forge Press}
}

@article{iaquinto2011creating,
  title={Creating communities of practice: scoping purposeful design},
  author={Iaquinto, Ben and Ison, Ray and Faggian, Robert},
  journal={Journal of Knowledge Management},
  volume={15},
  number={1},
  pages={4--21},
  year={2011},
  publisher={Emerald Group Publishing Limited}
}

@article{ma2019studying,
  title={Studying STEM faculty communities of practice through social network analysis},
  author={Ma, Shufeng and Herman, Geoffrey L and West, Matthew and Tomkin, Jonathan and Mestre, Jose},
  journal={The Journal of Higher Education},
  volume={90},
  number={5},
  pages={773--799},
  year={2019},
  publisher={Taylor \& Francis}
}

@article{cross2006using,
  title={Using social network analysis to improve communities of practice},
  author={Cross, Rob and Laseter, Tim and Parker, Andrew and Velasquez, Guillermo},
  journal={California Management Review},
  volume={49},
  number={1},
  pages={32--60},
  year={2006},
  publisher={SAGE Publications Sage CA: Los Angeles, CA}
}

@article{de2007investigating,
  title={Investigating patterns of interaction in networked learning and computer-supported collaborative learning: A role for Social Network Analysis},
  author={De Laat, Maarten and Lally, Vic and Lipponen, Lasse and Simons, Robert-Jan},
  journal={International Journal of Computer-Supported Collaborative Learning},
  volume={2},
  pages={87--103},
  year={2007},
  publisher={Springer}
}

@article{wellman2001computer,
  title={Computer networks as social networks},
  author={Wellman, Barry},
  journal={Science},
  volume={293},
  number={5537},
  pages={2031--2034},
  year={2001},
  publisher={American Association for the Advancement of Science}
}

@article{lin2016social,
  title={A social network analysis of teaching and research collaboration in a teachers' virtual learning community},
  author={Lin, Xiaofan and Hu, Xiaoyong and Hu, Qintai and Liu, Zhichun},
  journal={British Journal of Educational Technology},
  volume={47},
  number={2},
  pages={302--319},
  year={2016},
  publisher={Wiley Online Library}
}

@article{hatcher2022closeness,
  title={Closeness in a physics faculty online learning community predicts impacts in self-efficacy and teaching},
  author={Hatcher, Chase and Price, Edward and Smith, P Sean and Turpen, Chandra and Brewe, Eric},
  journal={arXiv preprint arXiv:2209.09306},
  year={2022}
}

@article{dou2019practitioner,
  title={Practitioner’s guide to social network analysis: Examining physics anxiety in an active-learning setting},
  author={Dou, Remy and Zwolak, Justyna P},
  journal={Physical Review Physics Education Research},
  volume={15},
  number={2},
  pages={020105},
  year={2019},
  publisher={APS}
}

@article{opsahl2009clustering,
  title={Clustering in weighted networks},
  author={Opsahl, Tore and Panzarasa, Pietro},
  journal={Social networks},
  volume={31},
  number={2},
  pages={155--163},
  year={2009},
  publisher={Elsevier}, 
  doi={https://doi.org/10.1016/j.socnet.2009.02.002 }
}

@article{fardet2021weighted,
  title={Weighted directed clustering: Interpretations and requirements for heterogeneous, inferred, and measured networks},
  author={Fardet, Tanguy and Levina, Anna},
  journal={Physical Review Research},
  volume={3},
  number={4},
  pages={043124},
  year={2021},
  publisher={APS}
}

@article{schank2005approximating,
  title={Approximating clustering coefficient and transitivity.},
  author={Schank, Thomas and Wagner, Dorothea},
  journal={Journal of Graph Algorithms and Applications},
  volume={9},
  number={2},
  pages={265--275},
  year={2005},
  publisher={Brown University, Providence, RI; University of Texas, Dallas}
}

@article{liu2014weighted,
  title={Weighted graph clustering for community detection of large social networks},
  author={Liu, Ruifang and Feng, Shan and Shi, Ruisheng and Guo, Wenbin},
  journal={Procedia Computer Science},
  volume={31},
  pages={85--94},
  year={2014},
  publisher={Elsevier}
}

@inproceedings{wang2015review,
  title={Review on community detection algorithms in social networks},
  author={Wang, Cuijuan and Tang, Wenzhong and Sun, Bo and Fang, Jing and Wang, Yanyang},
  booktitle={2015 IEEE international conference on progress in informatics and computing (PIC)},
  pages={551--555},
  year={2015},
  organization={IEEE}
}

@article{squartini2013reciprocity,
  title={Reciprocity of weighted networks},
  author={Squartini, Tiziano and Picciolo, Francesco and Ruzzenenti, Franco and Garlaschelli, Diego},
  journal={Scientific reports},
  volume={3},
  number={1},
  pages={2729},
  year={2013},
  publisher={Nature Publishing Group UK London}
}

@article{watts1998collective,
  title={Collective dynamics of ‘small-world’networks},
  author={Watts, Duncan J and Strogatz, Steven H},
  journal={nature},
  volume={393},
  number={6684},
  pages={440--442},
  year={1998},
  publisher={Nature Publishing Group}
}

@article{newman2003structure,
  title={The structure and function of complex networks},
  author={Newman, Mark EJ},
  journal={SIAM review},
  volume={45},
  number={2},
  pages={167--256},
  year={2003},
  publisher={SIAM}
}

@article{bavelas1950communication,
  title={Communication patterns in task-oriented groups},
  author={Bavelas, Alex},
  journal={The journal of the acoustical society of America},
  volume={22},
  number={6},
  pages={725--730},
  year={1950},
  publisher={AIP Publishing}
}

@article{freeman2002centrality,
  title={Centrality in social networks: Conceptual clarification},
  author={Freeman, Linton C},
  journal={Social network: critical concepts in sociology. Londres: Routledge},
  volume={1},
  pages={238--263},
  year={1978}
}

@article{opsahl2010nodecentrality,
  title={Node centrality in weighted networks: Generalizing degree and shortest paths},
  author={Opsahl, Tore and Agneessens, Filip and Skvoretz, John},
  journal={Social networks},
  volume={32},
  number={3},
  pages={245--251},
  year={2010},
  publisher={Elsevier}
}

@article{marion2016evaluating,
  title={Evaluating complex educational systems with quadratic assignment problem and exponential random graph model methods},
  author={Marion, Russ and Schreiber, Craig},
  journal={Complex dynamical systems in education: Concepts, methods and applications},
  pages={177--201},
  year={2016},
  publisher={Springer}
}

@article{maslov2004detection,
  title={Detection of topological patterns in complex networks: correlation profile of the internet},
  author={Maslov, Sergei and Sneppen, Kim and Zaliznyak, Alexei},
  journal={Physica A: Statistical Mechanics and its Applications},
  volume={333},
  pages={529--540},
  year={2004},
  publisher={Elsevier}
}

@article{hobson2021guide,
  title={A guide to choosing and implementing reference models for social network analysis},
  author={Hobson, Elizabeth A and Silk, Matthew J and Fefferman, Nina H and Larremore, Daniel B and Rombach, Puck and Shai, Saray and Pinter-Wollman, Noa},
  journal={Biological Reviews},
  volume={96},
  number={6},
  pages={2716--2734},
  year={2021},
  publisher={Wiley Online Library}
}

@article{croft2011hypothesis,
  title={Hypothesis testing in animal social networks},
  author={Croft, Darren P and Madden, Joah R and Franks, Daniel W and James, Richard},
  journal={Trends in ecology \& evolution},
  volume={26},
  number={10},
  pages={502--507},
  year={2011},
  publisher={Elsevier}
}

@article{cranmer2017navigating,
  title={Navigating the range of statistical tools for inferential network analysis},
  author={Cranmer, Skyler J and Leifeld, Philip and McClurg, Scott D and Rolfe, Meredith},
  journal={American Journal of Political Science},
  volume={61},
  number={1},
  pages={237--251},
  year={2017},
  publisher={Wiley Online Library}
}

@article{newman2008physics,
  title={The physics of networks},
  author={Newman, Mark E J},
  journal={Physics today},
  volume={61},
  number={11},
  pages={33--38},
  year={2008},
  publisher={AIP Publishing}
}

@article{newman2001random,
  title={Random graphs with arbitrary degree distributions and their applications},
  author={Newman, Mark E J and Strogatz, Steven H. and Watts, Duncan J.},
  journal={Physical review E},
  volume={64},
  number={2},
  pages={026118},
  year={2001},
  publisher={APS}
}

@article{sun2014understanding,
  title={Understanding lurkers in online communities: A literature review},
  author={Sun, Na and Rau, Patrick Pei-Luen and Ma, Liang},
  journal={Computers in Human Behavior},
  volume={38},
  pages={110--117},
  year={2014},
  publisher={Elsevier}
}

@misc{Slack, 
  author = {Butterfield, S. and Costello, E. and Henderson, C. and Mourachov, S.},
  title={Slack is your productivity platform}, 
  url={https://slack.com/}, 
  journal={Slack Technologies, San Francisco, California}, 
	year={2009}}

@book{kadushin2012understanding,
  title={Understanding social networks: Theories, concepts, and findings},
  author={Kadushin, Charles},
  year={2012},
  publisher={Oxford university press}
}

@book{borgatti2018analyzing,
  title={Analyzing social networks},
  author={Borgatti, Stephen P and Everett, Martin G and Johnson, Jeffrey C},
  year={2018},
  publisher={Sage: Thousand Oaks, California }
}

@book{newman2010networks,
  title={Networks: An Introduction},
  author={Newman, Mark},
  year={2010},
  publisher={Oxford university press}
}

@book{fornito2016fundamentals,
  title={Fundamentals of brain network analysis},
  author={Fornito, Alex and Zalesky, Andrew and Bullmore, Edward},
  year={2016},
  publisher={Academic press}
}

@misc{hanneman2005introduction,
  title={Introduction to social network methods},
  author={Hanneman, Robert A and Riddle, Mark},
  year={2005},
  publisher={University of California, Riverside, California}
}

@book{wasserman1994social,
  title={Social network analysis: Methods and applications},
  author={Wasserman, Stanley},
  publisher={The Press Syndicate of the University of Cambridge, Cambridge, United Kingdom},
  year={1994}
}

@manual{R,
     title = {R: A Language and Environment for Statistical Computing},
     author = {{R Core Team}},
     organization = {R Foundation for Statistical Computing},
     address = {Vienna, Austria},
    year = {2021},
    url = {https://www.R-project.org/},
}

@article{igraph1,
    title = {The igraph software package for complex network research},
    author = {Gabor Csardi and Tamas Nepusz},
    journal = {InterJournal},
    volume = {Complex Systems},
    pages = {1695},
    year = {2006},
    url = {https://cran.r-project.org/web/packages/igraph/citation.html},
}

@manual{igraph2,
    title = {{igraph}: Network Analysis and Visualization in R},
    author = {Gábor Csárdi and Tamás Nepusz and Vincent Traag and Szabolcs Horvát and Fabio Zanini and Daniel Noom and Kirill Müller},
    year = {2025},
    note = {R package version 2.1.4},
    doi = {10.5281/zenodo.7682609},
    url = {https://doi.org/10.5281/zenodo.7682609},
}

@book{tnet,
    title = {Structure and Evolution of Weighted Networks},
    author = {Tore Opsahl},
    year = {2009},
    publisher = {University of London (Queen Mary College), London,
      UK},
    pages = {104-122},
    url = {http://toreopsahl.com/publications/thesis/},
}

@article{kk,
  title={An algorithm for drawing general undirected graphs},
  author={Kamada, Tomihisa and Kawai, Satoru and others},
  journal={Information processing letters},
  volume={31},
  number={1},
  pages={7--15},
  year={1989},
  publisher={Citeseer}
}

@article{newmanmej2003whysocialnetworksaredifferent,
  title={Whysocialnetworksaredifferent fromothertypesofnetworks},
  author={Newman M E J, Park J},
  journal={PhysicalReviewE},
  volume={68},
  number={3},
  pages={036122},
  year={2003}
}

@book{tagarelli2018mining,
  title={Mining lurkers in online social networks: Principles, models, and computational methods},
  author={Tagarelli, Andrea and Interdonato, Roberto},
  year={2018},
  publisher={Springer}
}

@article{preece2004top,
  title={The top five reasons for lurking: improving community experiences for everyone},
  author={Preece, Jenny and Nonnecke, Blair and Andrews, Dorine},
  journal={Computers in human behavior},
  volume={20},
  number={2},
  pages={201--223},
  year={2004},
  publisher={Elsevier}
}

@article{elliott2017evolution,
  title={The evolution from traditional to online professional development: A review},
  author={Elliott, Joshua C},
  journal={Journal of Digital Learning in Teacher Education},
  volume={33},
  number={3},
  pages={114--125},
  year={2017},
  publisher={Taylor \& Francis}
}

@article{abdullah2016equity,
  title={Equity and inclusion in online community forums: an interview with Steven Clift},
  author={Abdullah, Carolyne and Karpowitz, Christopher F and Raphael, Chad},
  journal={Journal of Deliberative Democracy},
  volume={12},
  number={2},
  year={2016},
  pages={1-14},
  publisher={University of Westminster Press}
}

@article{mchugh2020uncovering,
  title={Uncovering themes in personalized learning: Using natural language processing to analyze school interviews},
  author={McHugh, David and Shaw, Sarah and Moore, Travis R and Ye, Leafia Zi and Romero-Masters, Philip and Halverson, Richard},
  journal={Journal of Research on Technology in Education},
  volume={52},
  number={3},
  pages={391--402},
  year={2020},
  publisher={Taylor \& Francis}
}

@article{WangWang2022,
  author = {Wang, Dakuo and Wang, Haoyu and Yu, Mo and Ashktorab, Zahra and Tan, Ming},
  title = {Group Chat Ecology in Enterprise Instant Messaging: How Employees Collaborate Through Multi-User Chat Channels on Slack},
  year = {2022},
  issue_date = {April 2022},
  publisher = {Association for Computing Machinery},
  address = {New York, NY, USA},
  volume = {6},
  doi = {10.1145/3512941},
  journal = {Proc. ACM Hum.-Comput. Interact.},
  articleno = {94},
  numpages = {14}
}

@article{COSTASILVA2022111397,
title = {A qualitative analysis of themes in instant messaging communication of software developers},
journal = {Journal of Systems and Software},
volume = {192},
pages = {111397},
year = {2022},
issn = {0164-1212},
doi = {https://doi.org/10.1016/j.jss.2022.111397},
author = {Camila {Costa Silva} and Matthias Galster and Fabian Gilson}
}

@article{martinez2003combining,
  title={Combining qualitative evaluation and social network analysis for the study of classroom social interactions},
  author={Mart{\i}nez, A and Dimitriadis, Yannis and Rubia, Bartolom{\'e} and G{\'o}mez, Eduardo and De la Fuente, Pablo},
  journal={Computers \& education},
  volume={41},
  number={4},
  pages={353--368},
  year={2003},
  publisher={Elsevier}
}

@article{Singh2025,
author = {Singh, Shashank Sheshar and Muhuri, Samya and Kumar, Sumit and Barua, Jayendra},
title = {From Nodes to Knowledge: Exploring Social Network Analysis in Education},
year = {2025},
issue_date = {February 2025},
publisher = {Association for Computing Machinery},
address = {New York, NY, USA},
volume = {19},
number = {1},
issn = {1559-1131},
url = {https://doi-org.proxy2.cl.msu.edu/10.1145/3707463},
doi = {10.1145/3707463},
journal = {ACM Trans. Web},
month = jan,
articleno = {7},
pages = {1-36}
}

@article{wilson2009user,
  title={Beyond social graphs: User interactions in online social networks and their implications},
  author={Wilson, Christo and Sala, Alessandra and Puttaswamy, Krishna PN and Zhao, Ben Y},
  journal={ACM Transactions on the Web (TWEB)},
  volume={6},
  number={4},
  pages={1--31},
  year={2012},
  publisher={ACM New York, NY, USA}
}

@misc{egbolger,
	author = {egbolger},
	doi = {10.5281/zenodo.12574849},
	month = jun,
	publisher = {Zenodo},
	title = {egbolger/PICUP\_Publication\_Codeshare: Initial Release},
	url = {10.5281/zenodo.12574849},
	version = {v1.0},
	year = 2025}

\newpage
{
\appendix
\section{Network Representations for Other Channels} \label{appendix1}

In Figures~\ref{fig:dwc_advmech}-\ref{fig:dwc_glow}, we show the network representations for the Advanced Mechanics, Upper Mid-West, Why We Do This, Trinket, and Glowscript channels as described in Section~\ref{sec:other}. We notice the coding focused channels (Trinket, Glowscript, and previously discussed, Jupyter) have more interactions than other channels in the Slack Workspace.

\begin{figure}[htp]
    \centering
    \includegraphics[width=0.45\textwidth]{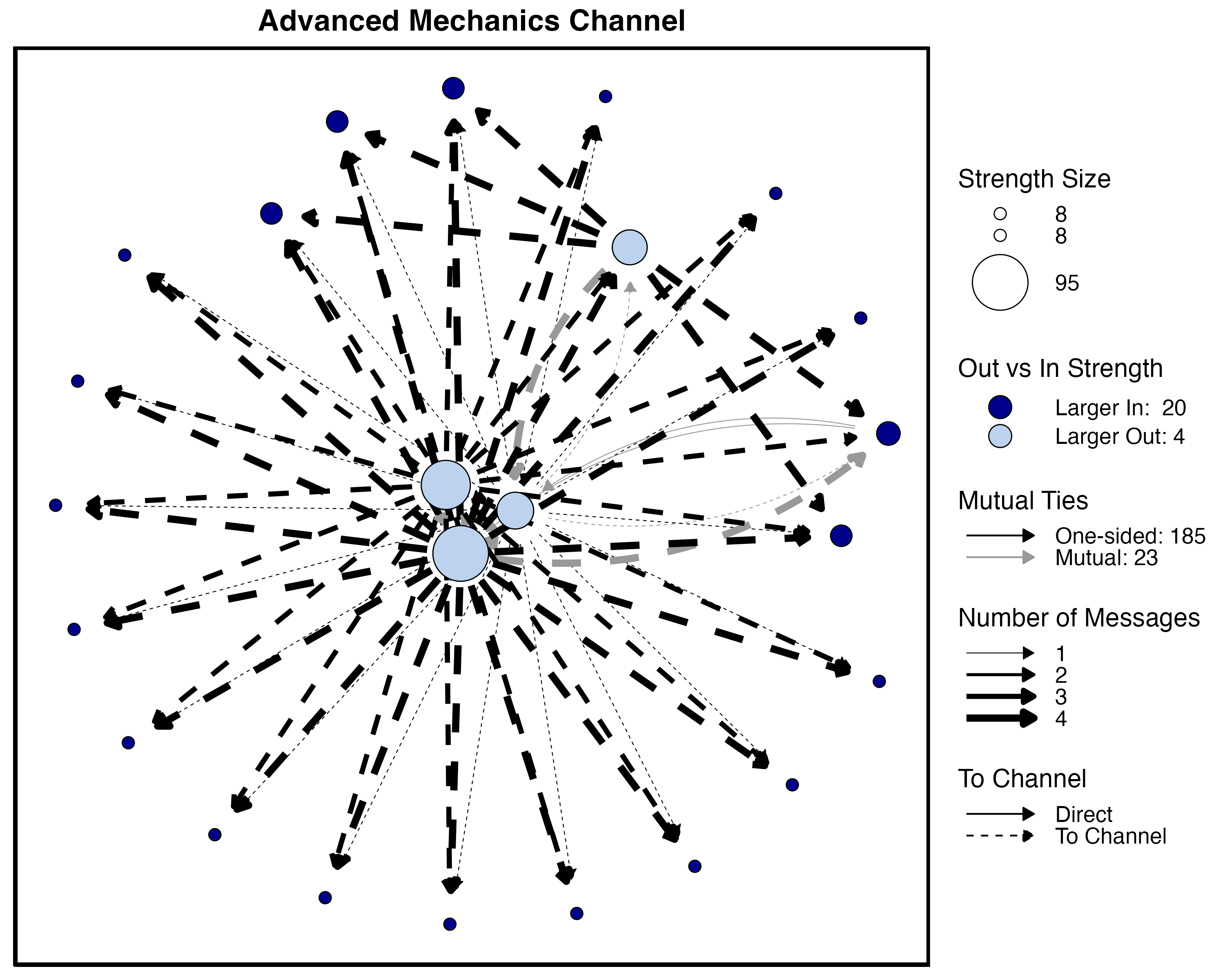}
    \caption{Network Representation for the Advanced Mechanics channel. The channel is similar in size to the Advanced Thermodynamics channel and similarly has a focus on curriculum of Advanced Mechanics.  }
    \label{fig:dwc_advmech}
\end{figure}

\begin{figure}[htp]
    \centering
    \includegraphics[width=0.45\textwidth]{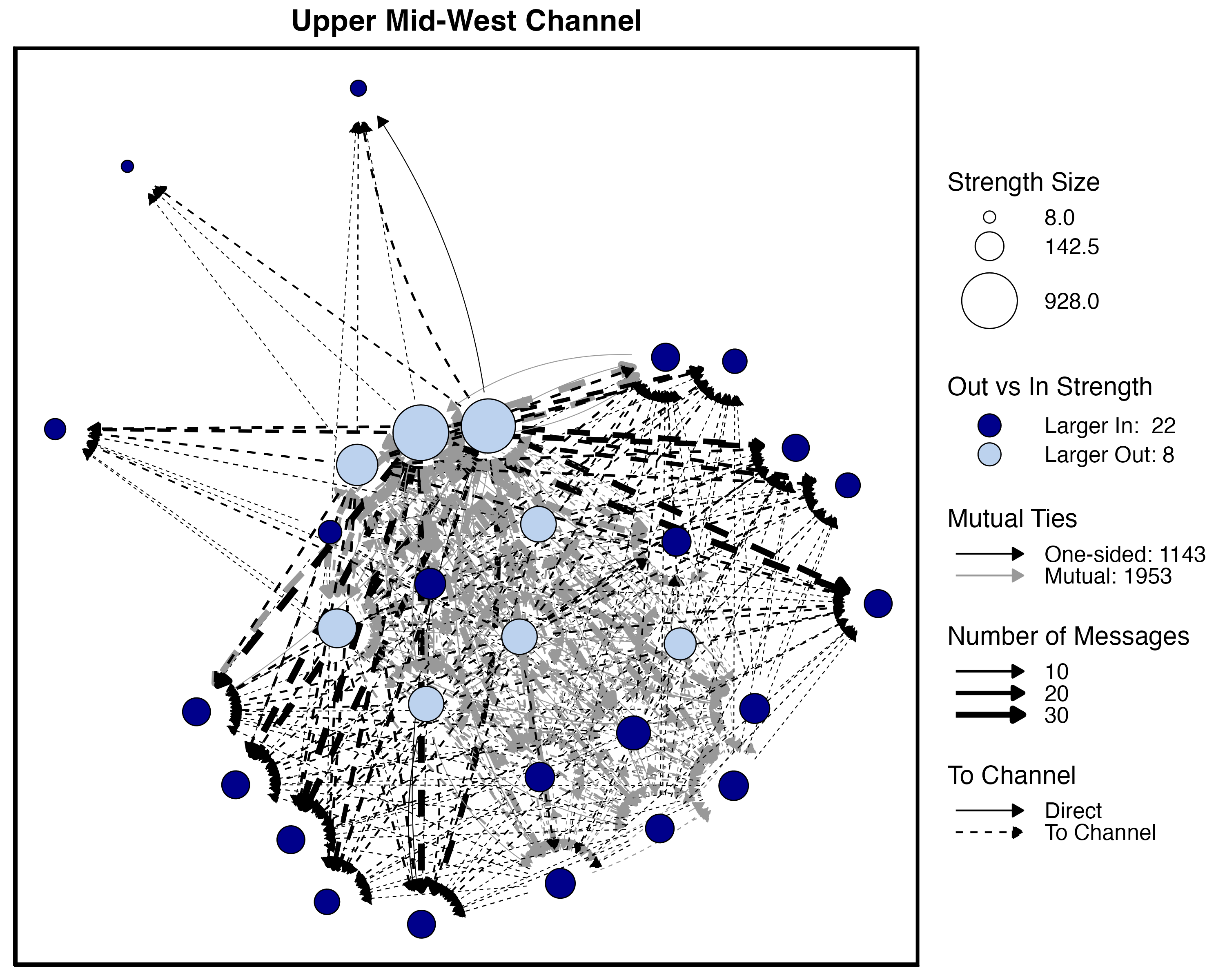}
    \caption{Network Representation for the Upper Mid-West channel. The channel is one of a few that are regionally focused.}
    \label{fig:dwc_umw}
\end{figure}

\begin{figure}[htp]
    \centering
    \includegraphics[width=0.35\textwidth]{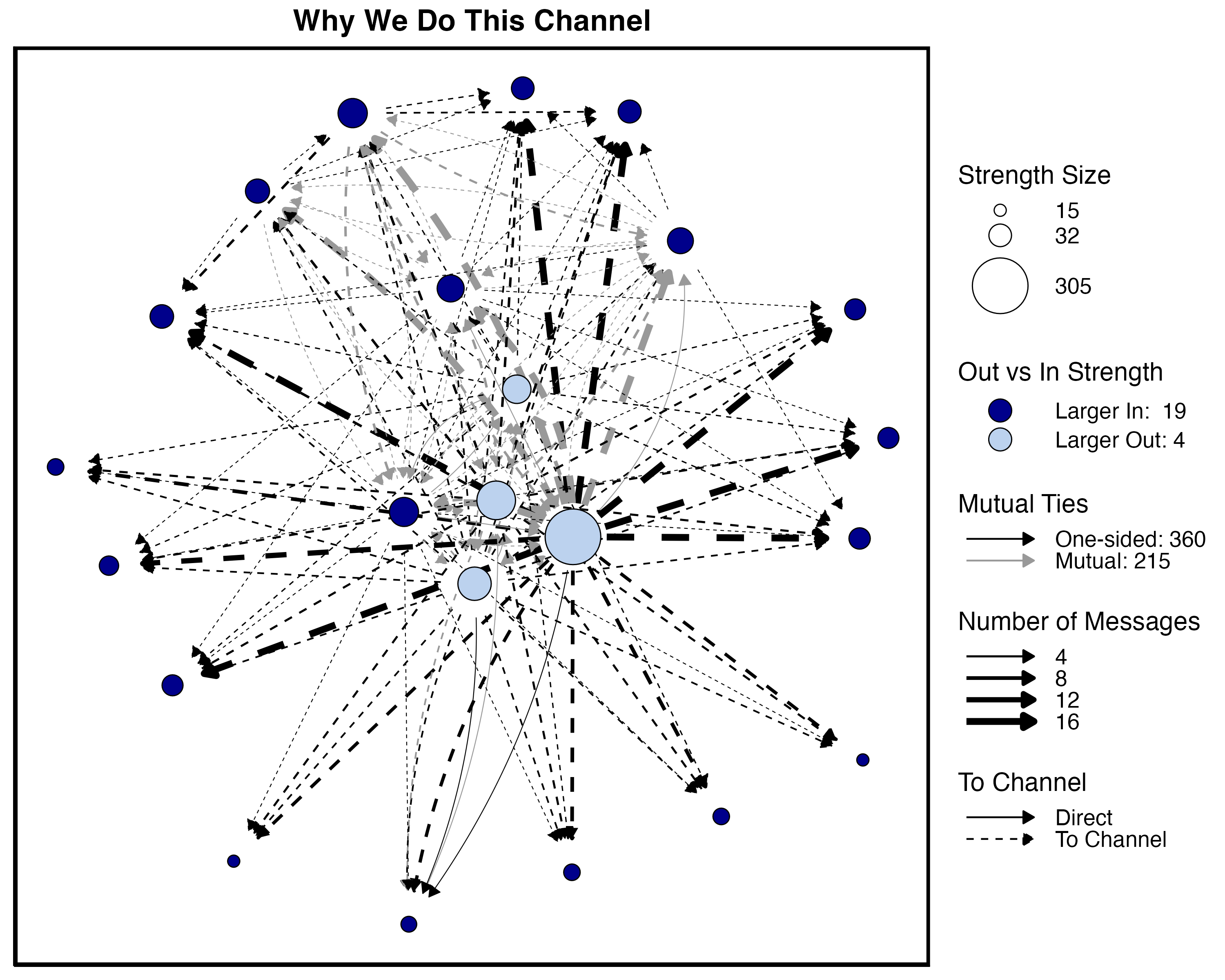}
    \caption{Network Representation for the Why We Do This channel. In the channel, users discuss their motivation for educating students and participating in this community. }
    \label{fig:dwc_wwdt}
\end{figure}

\begin{figure}[htp]
    \centering
    \includegraphics[width=0.35\textwidth]{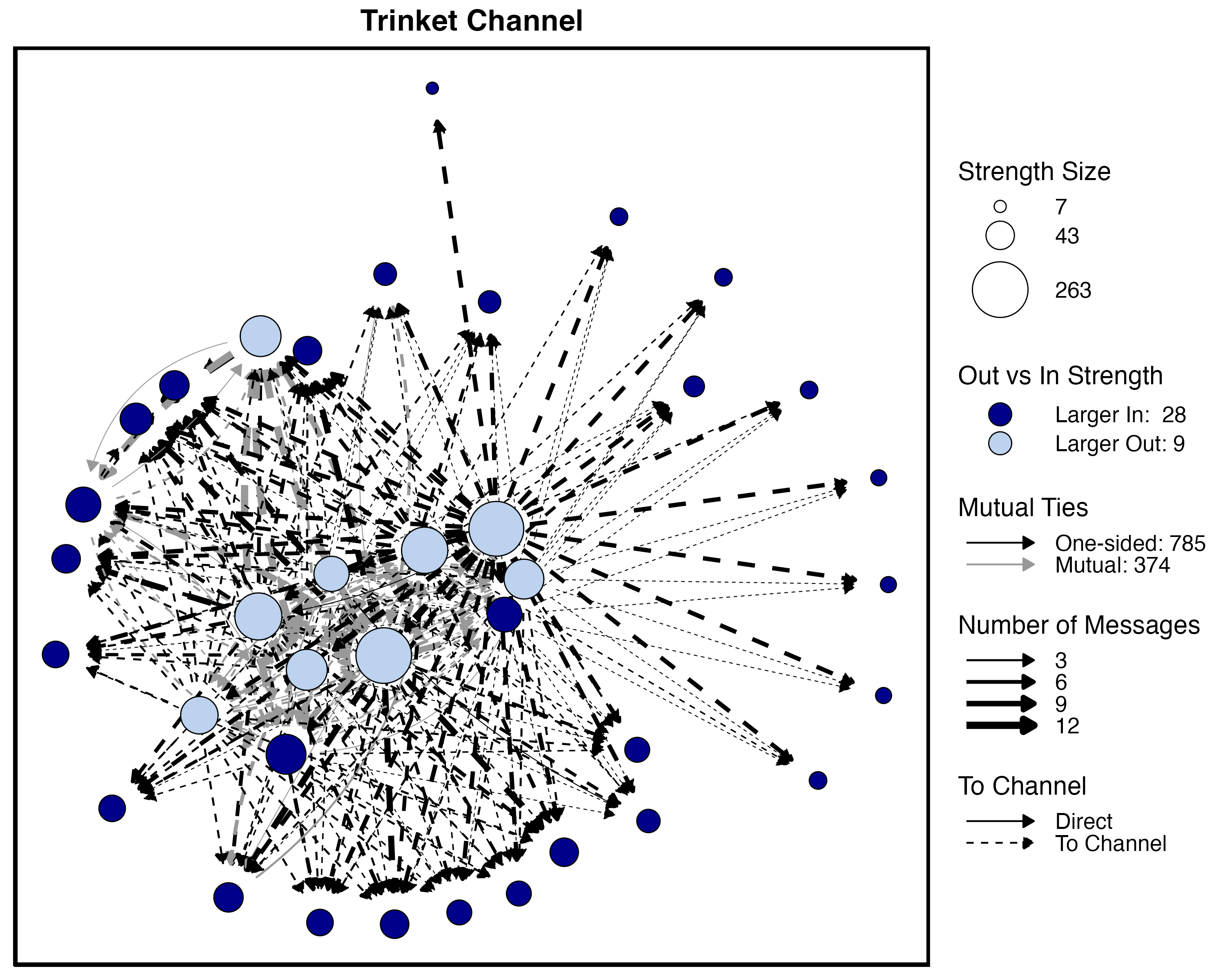}
    \caption{Network Representation for the Trinket channel. Trinket is another coding tool allowing users to run Python code interactively in a web browser.}
    \label{fig:dwc_trin}
\end{figure}

\begin{figure}[htp]
    \centering
    \includegraphics[width=0.35\textwidth]{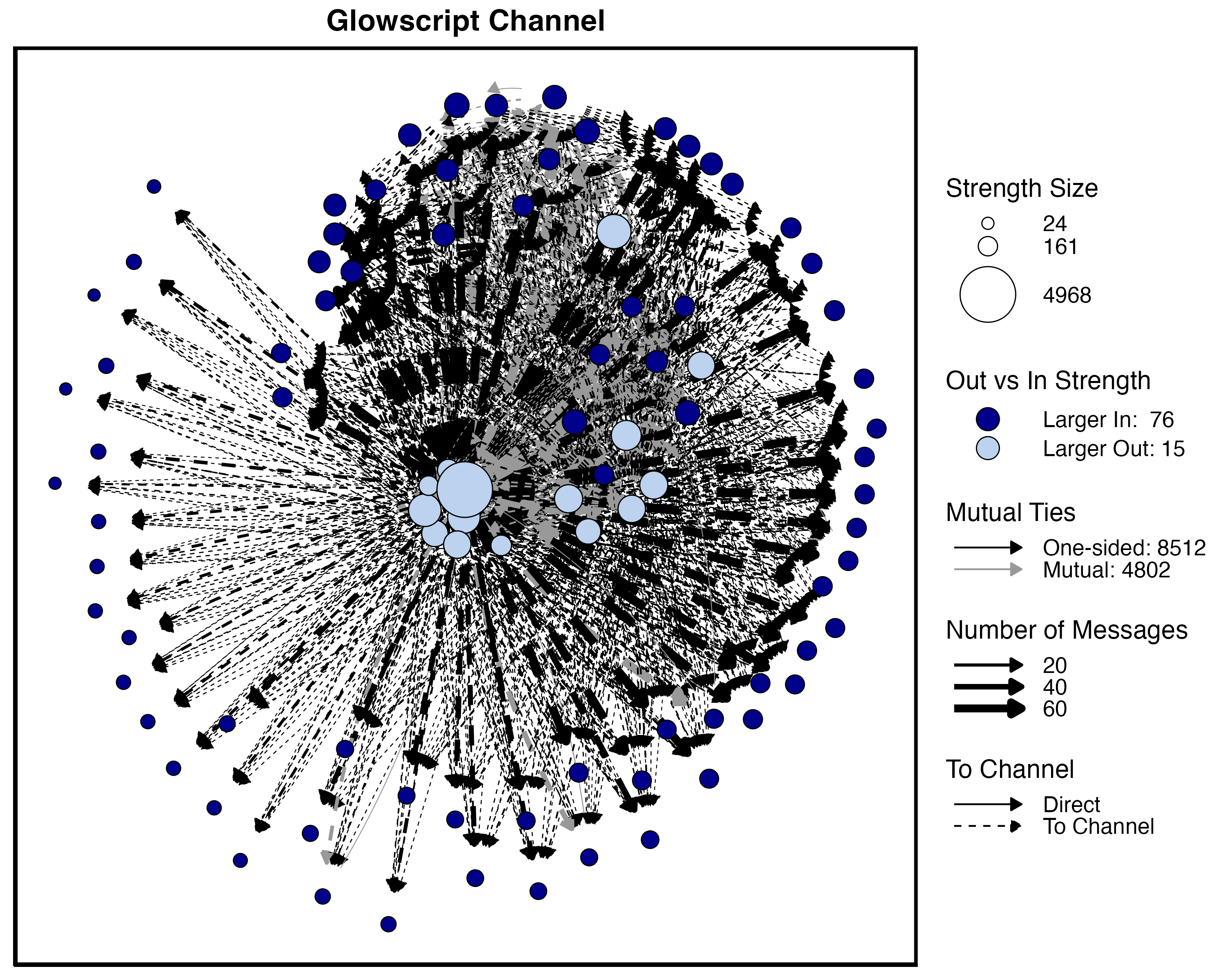}
    \caption{Network Representation for the Glowscript channel. Glowscript is a tool for creating visualizations in Python.}
    \label{fig:dwc_glow}
\end{figure}

\clearpage
\newpage
\section{Configuration Model Walkthrough} \label{appendix2}
\begin{figure}[H]
    \centering
    \begin{subfigure}[b]{0.38\textwidth}
        \centering
         \includegraphics[width=\textwidth]{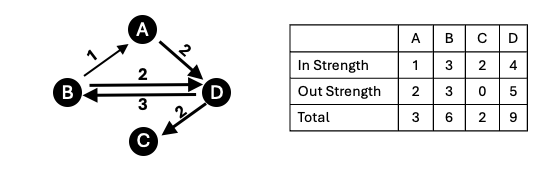}
        \caption{} 
        \label{fig:configex}
        \end{subfigure}
    \begin{subfigure}[b]{0.38\textwidth}
       \centering
       \includegraphics[width=\textwidth]{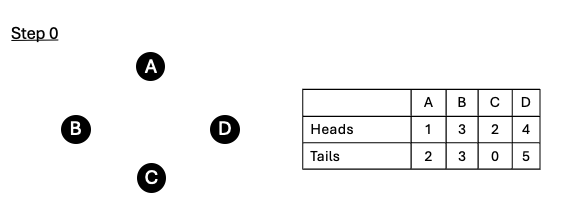}
        \caption{ } 
        \label{fig:configex0}
    \end{subfigure}
    \begin{subfigure}[b]{0.38\textwidth}
       \centering
      \includegraphics[width=\textwidth]{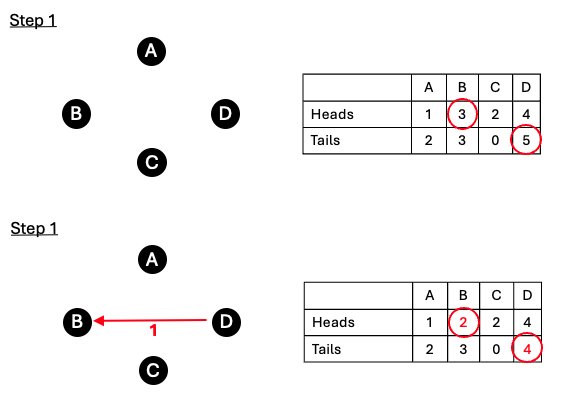}
        \caption{ } 
        \label{fig:configex1}
    \end{subfigure}
    \begin{subfigure}[b]{0.38\textwidth}
       \centering
      \includegraphics[width=\textwidth]{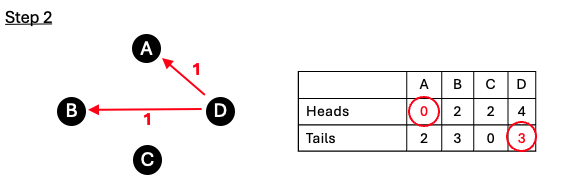}
        \caption{ } 
        \label{fig:configex2}
    \end{subfigure}
     \begin{subfigure}[b]{0.38\textwidth}
       \centering
      \includegraphics[width=\textwidth]{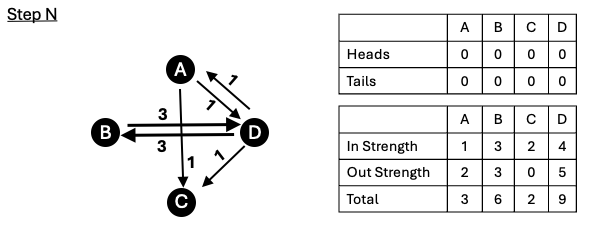}
        \caption{ } 
        \label{fig:configexN}
    \end{subfigure}
       \caption{A walkthrough of the first couple steps of the Configuration Model on an example network. Figure~\ref{fig:configex} shows an example network with the associated in-strength and out-strength values. Figures~\ref{fig:configex0}-\ref{fig:configex2} show the first couple iterations of the randomization method. Figure~\ref{fig:configexN} shows an example of randomized network preserving in and out strengths. }
       \label{fig:configex_all}
 \end{figure}

\begin{algorithm}[H]
  \caption{Configuration Model}
  \label{alg:config}
   \begin{algorithmic}[1]
    \State heads $ \gets [s^{in}_1, s^{in}_2, \dots, s^{in}_n]$, where $n = $ number of nodes
    \State tails $ \gets [s^{out}_1, s^{out}_2, \dots, s^{out}_n]$, where $n = $ number of nodes \\
    
    \While{sum(heads) $ > 0$}\\

    
        \State head\_index$=$random index s.t. heads[head\_index] $\neq 0$
        \State tail\_index$=$random index s.t. tails[tail\_index] $\neq 0$\\



    \State heads[head\_index] $\gets$ heads[head\_index] - 1
    \State tails[tail\_index] $\gets$ tails[tail\_index] - 1 \\

    \If{edge between head\_index and tail\_index exists}
        \State increase edge weight
    \Else
        \State add edge between head\_index and tail\_index
    \EndIf
    \EndWhile
   \end{algorithmic}
\end{algorithm}

In this Section, we describe the steps of the Configuration Randomization Method \citep{newman2001random, newman2003structure, newman2010networks} through an example. The algorithm in pseudocode is presented in Algorithm~\ref{alg:config}.

Figure~\ref{fig:configex} depicts an example network with 4 nodes (A, B, C, D) and a set of edges connecting them. The table on the right contains the calculated values for the associated in strength and out strength of the nodes in the network. As mentioned, the Configuration Model preserves the strength distribution of the nodes in the network. For the sake of the generation method, we create a copy of the in and out strength lists, renaming that ``heads'' and ``tails.'' 

As in Figure~\ref{fig:configex0}, at the beginning of the algorithm, we start with the same node set as the original network. We then randomly select an index from the ``heads'' list and an index from the ``tails'' list. Here, Node B is chosen at the head node and Node D is chosen as the tail node. Essentially, we are randomly picking two nodes to add an edge between, with the assumption that the in/out strength values are non-zero. We decrement the values associated to the chosen head and tail indices. Then, we add an edge from the chosen tail index to the chosen head index, creating a new edge if one does not already exist or simply incrementing the edge weight. This process is shown in Figure~\ref{fig:configex1}, where we add an edge from node D to node B with edge weight $1$.

In Figure~\ref{fig:configex2}, we show another iteration of the Configuration Model. In this iteration, Node A is selected as the head node and Node D is selected as the tail node. In the same way, we decrement the values at those indices in the ``heads'' and ``tails'' lists and add the new edge.  

We repeat this process of randomly selecting nodes to add an edge between until there are no more available edges - meaning the sum of the in/out strengths list is $0$. An example of a completed output from this randomization method can be seen in Figure~\ref{fig:configexN}. As shown in the tables to the right, the ``heads'' and ``tails'' lists are zeroed out, the in strength and out strength lists are identical to the original example network, and the network on the left has a different configuration than the example. 

\clearpage
\newpage
\section{Addressing Repeats Networks in the Configuration Model} \label{appendix3}

As mentioned in Section~\ref{sec:config}, when implementing the configuration model, there is an upper limit on the number of possible network configurations. Repeated network configurations can affect the distribution of metrics values and thus, skew the resulting probability value. To address the potential impact of repeated configurations on our analysis results, we conducted a simulation to quantify the number of identical networks in the random ensemble. 

In the configuration model, the in/out strength associated to each node are fixed. Keeping this in mind, we only quantify the number of identical graphs, not particularly isomorphic graphs. We conducted three simulations: (1) a directed 4-cycle with edge weight 1, (2) a directed 10-cycle with edge weight 1, (3) the Advanced Thermodynamics network. 

For the directed 4-cycle, there are 9 unique configurations preserving the in/out strength distributions, which is achieved with a small number of iterations (e.g. 25 iterations). Increasing to a 10-cycle, we generated ensembles of increasing sizes from 100 to 10,000. As seen in Table~\ref{tab:unique} and Figure~\ref{fig:unique}, it takes an ensemble size of 2,000 to see repeats of identical networks in the ensemble, where 0.25\% percent of repeats. Even for a large ensemble size of 10,000, the percentage of repeats is very low at 0.38\%.    

\begin{table*}[htp] 
    \centering
    \caption{The number of network configurations in the ensemble with the associated number of unique networks for a directed 10-cycle}
    \label{tab:unique}
    \begin{tabularx}{\linewidth}{c|c|c}\hline\hline
        Total Number in Ensemble  & Number of Unique Networks  &  Percentage of Repeats \\\hline
        100 & 100 & 0 \%\\ 
        250 & 250 & 0 \%\\ 
        500 & 500 & 0 \%\\ 
        750 & 750 & 0 \%\\ 
        1000 & 1000 & 0 \%\\ 
        2000 & 1995 & 0.25\% \\ 
        3000 & 2996 & 0.13\% \\ 
        4000 & 3992 & 0.20\% \\ 
        5000 & 4992 & 0.16\% \\ 
        7500 & 7489 & 0.15\%\\ 
        10000 & 9962 & 0.38\% \\ \hline\hline
    \end{tabularx}
\end{table*}

\begin{figure}[htp]
    \centering
    \includegraphics[width=0.4\textwidth]{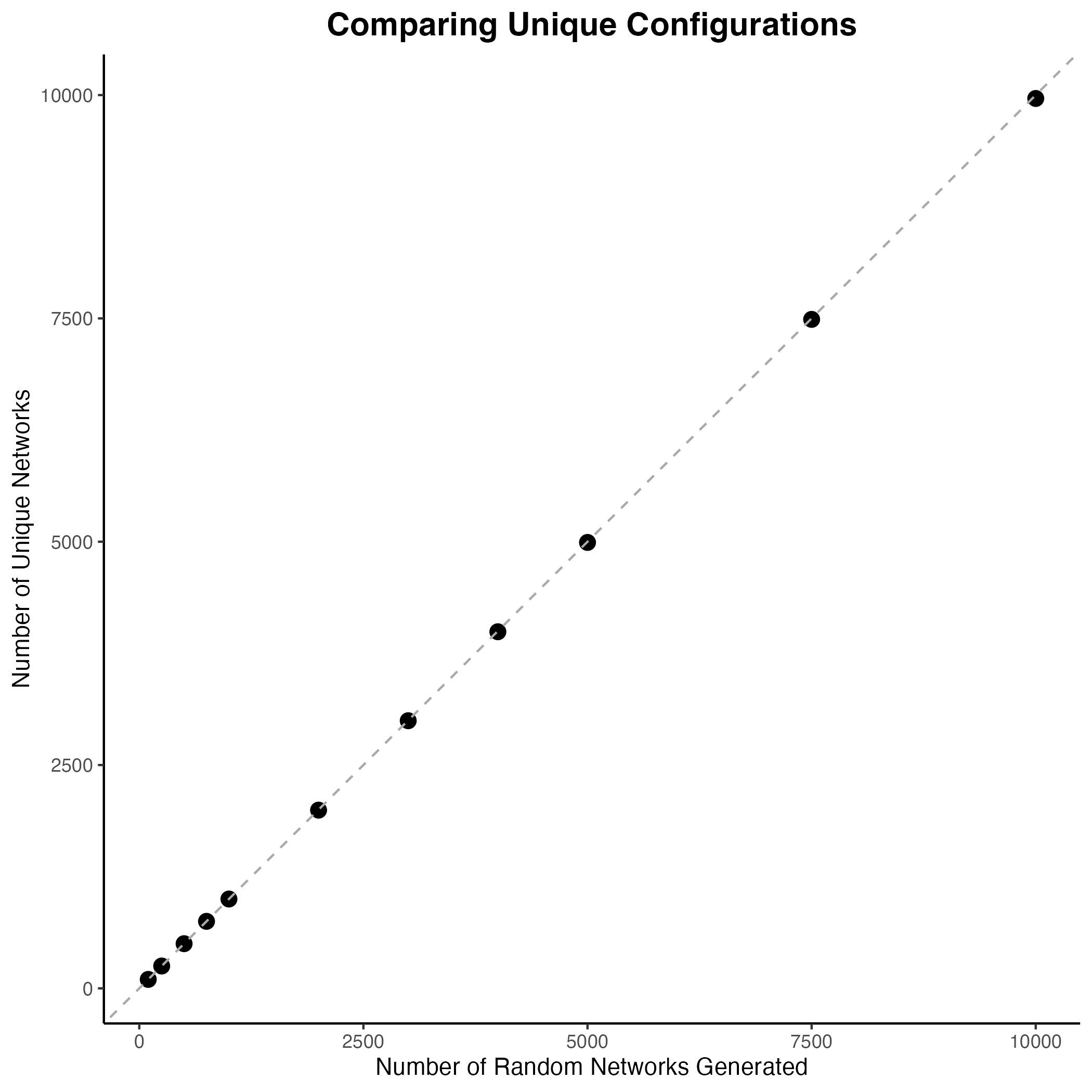}
    \caption{Visualizing the number of networks in the ensemble with the number of unique configurations in the ensemble. The same information as Table~\ref{tab:unique}. A nearly perfect identity line, with slight deviation at the higher values. }
    \label{fig:unique}
\end{figure}

Generating an ensemble of 10,000 networks for the Advanced Thermodynamics network did not result in any identical repeats. Given that this network has more nodes, more edges, and higher weighted edges than the directed 10-cycle, this is not necessarily surprising. Thus, the likelihood of identical repeats in our ensemble for the Advanced Thermodynamics network - and subsequently, for our other networks that are larger in node and edge count - is very small. Thus, we are not overly concerned with the potential impact of identical repeats in our ensembles generated with the configuration model, and hence with the result that follows.



\end{document}